\documentclass[11pt,a4paper]{article}
\usepackage{jheppub}
\usepackage[T1]{fontenc}
\usepackage{amssymb}
\usepackage{mathtools}
\usepackage{xcolor}

\usepackage{hyperref}
\usepackage{epsfig}
\usepackage{amsmath,latexsym,amssymb}
\usepackage{graphicx}
\usepackage{braket}
\usepackage{ulem}
\usepackage[noabbrev]{cleveref}

\usepackage{mathrsfs}

\makeatletter
\newcommand{\doublewidetilde}[1]{{%
  \mathpalette\double@widetilde{#1}%
}}
\newcommand{\double@widetilde}[2]{%
  \sbox\z@{$\m@th#1\widetilde{#2}$}%
  \ht\z@=.9\ht\z@
  \widetilde{\box\z@}%
}
\makeatother

\def\be{\begin{equation}}
\def\ee{\end{equation}}
\def\ba{\begin{eqnarray}}
\def\ea{\end{eqnarray}}

\newcommand{\comment}[1]{}

\newcommand{\eea}{\end{eqnarray}}

\author{
Iustin Surubaru${}^{1}$,\, Bin Zhu${}^{1}$\\[0.5cm]

$^1${\it School of Mathematics and Maxwell Institute for Mathematical Sciences,\\ University of Edinburgh,
EH9 3FD, U.K. }\\[0.2cm]
}

\emailAdd{iustin.surubaru@ed.ac.uk}
\emailAdd{bzhu@ed.ac.uk}
\title{Carrollian Amplitudes and Holographic Correlators in AdS$_3$/CFT$_2$}

\abstract{We study Carrollian amplitudes of massless scalars in (1+2) Minkowski space. Using the prescription recently shown by Alday et al. \cite{Alday:2024yyj} originally designed for the AdS$_4$ Witten diagrams, we show that AdS$_3$ Witten diagrams in position space in the flat space limit reduce to Carrollian amplitudes. The flat space limit in the bulk is implemented by the Carrollian limit at the boundary. Focusing on four-point correlators with contact and exchange diagrams, we show that the Carrollian limit makes the universality of the bulk point singularity manifest upon performing analytic continuation to the Lorentzian signature of the boundary correlators. Unlike four-point Carrollian amplitudes in (1+3) dimensions, the (1+2) dimensional ones are non-distributional, having analytic properties simpler than the AdS correlators. We also observe for the first time a double copy structure of Carrollian amplitudes.}

\makeatletter
\gdef\@fpheader{}
\makeatother

\begin{document}
\maketitle

\section{Introduction} \label{sec:1}

Flat holography has gained a lot of interest in recent years. Two seemingly different proposals of flat holography have emerged: Celestial holography and Carrollian holography. The main conjecture of celestial holography is that the dual theory of quantum gravity in $d$-dimensional asymptotically flat space is a celestial CFT living on the $(d-2)$-dimensional celestial sphere at null infinity\footnote{See e.g. \cite{Raclariu:2021zjz} for reviews.}. In contrast, the main proposal of Carrollian holography is that the dual theory is a $(d-1)$ dimensional Carrollian CFT living at null infinity \cite{Donnay:2022aba,Bagchi:2022emh,Donnay:2022wvx,Banerjee:2018gce,Banerjee:2019prz}. The Carrollian framework is reached in the limit in which the speed of light vanishes, i.e. $c\rightarrow0$ \cite{Barnich:2012rz,Duval:2014uva,Bagchi:2019xfx,Henneaux:2021yzg,deBoer:2021jej,Chen:2023pqf,Alday:2024yyj}. In both proposals, the $S$-matrix of the bulk theory is dually recovered through the correlators of the appropriate CFT living either on the whole of null infinity or the celestial sphere. It has been shown that from the perspective of scattering amplitudes, celestial and Carrollian holography are closely related by a change of basis of scattering amplitudes \cite{Donnay:2022aba,Bagchi:2022emh,Donnay:2022wvx,Banerjee:2018gce,Banerjee:2019prz}.

When expressed in position space at null infinity, flat space amplitudes are mapped to Carrollian amplitudes which behave as correlation functions in a Carrollian CFT. Most of the work on Carrollian amplitudes has focused on massless scattering in four dimensions, see e.g. \cite{Donnay:2022aba,Bagchi:2022emh,Donnay:2022wvx,Salzer:2023jqv,Saha:2023abr,Nguyen:2023vfz,Nguyen:2023miw,Mason:2023mti,Bagchi:2023cen,Liu:2024nfc,Have:2024dff,Stieberger:2024shv,Adamo:2024mqn,Alday:2024yyj,Banerjee:2024hvb,Liu:2024llk,Ruzziconi:2024zkr,Ruzziconi:2024kzo,Chakrabortty:2024bvm, Nguyen:2025sqk}. In this work, we initiate a study of Carrollian amplitudes associated with massless scalar scattering in three dimensions. This includes massless scalars with self cubic and quartic interactions, and massless scalars coupled to Chern-Simons gauge fields and gravity.

The codimension-$1$ feature suggests that Carrollian holography would be more naturally related to the flat space limit of the AdS/CFT correspondence \cite{Maldacena:1997re, Witten:1998qj, Aharony:1999ti} which is by far the most successful realization of the holographic principle. Understanding the flat space limit of AdS/CFT has been a long-standing effort with different proposals \cite{Susskind:1998vk, Polchinski:1999ry, Giddings:1999jq,Marotta:2024sce,Gary:2009ae, Penedones:2010ue,Fitzpatrick:2011hu,Alday:2017vkk,Hijano:2019qmi}. See, e.g. \cite{Li:2021snj} for a comprehensive discussion and references therein. 

With the new notions of flat holography introduced in recent years, there has been increased interest in relating the new formalisms to the flat space limit of AdS/CFT \cite{Lam:2017ofc,deGioia:2022fcn,deGioia:2024yne, Bagchi:2023cen, Bagchi:2023fbj, Alday:2024yyj}. In particular, the authors of \cite{Alday:2024yyj} gave a prescription for how to reproduce Carrollian amplitudes of massless particles in four dimensions by taking the Carrollian limit of holographic correlators in AdS$_4$/CFT$_3$. The flat space limit in the bulk is realized by the Carrollian limit at the boundary. In the usual set-up of taking the flat limit of AdS, the boundary operators are restricted to strips, a consequence of their dual bulk description. However, the advantage of the prescription in \cite{Alday:2024yyj} is that the boundary insertions are not restricted a priori, which is more natural from an intrinsic boundary perspective. It turns out that the analytic continuation from Euclidean to Lorentzian signature is crucial for taking the Carrollian limit at the boundary. The presence of a bulk point singularity arises as a consequence of the Carrollian limit. In this work, we apply their prescription to holographic correlators in AdS$_3$/CFT$_2$ and confirm that the matching also holds in this case. This provides strong evidence that Carrollian holography is a natural candidate related to the flat space limit of AdS/CFT.

In parallel, scattering amplitudes in gravity and Yang-Mills theory exhibit a remarkable double-copy structure \cite{Kawai:1985xq,Bern:2008qj}, known as the Kawai-Lewellen-Tye (KLT) relations \cite{Kawai:1985xq} and the Bern-Carrasco-Johansson (BCJ) double copy \cite{Bern:2008qj}. The double copy relations state that gravitational amplitudes
can be obtained by a well-defined “squaring” of gauge theory amplitudes. The double copy relations have been well studied in momentum eigenstate basis with a rich mathematical structure and a wide range of applications. See e.g. \cite{Bern:2019prr,Adamo:2022dcm} for reviews. As for other bases, e.g, boost eigenstate basis \cite{Arkani-Hamed:2020gyp,Stieberger:2018onx}, the double-copy structure of celestial amplitudes has been studied in \cite{Stieberger:2018onx,Fan:2020xjj,Casali:2020vuy,Casali:2020uvr, Surubaru:2025qhs}. Surprisingly, in this work we find an interesting double copy structure of Carrollian amplitudes.

The paper is organized as follows. In section \ref{sec:2}, we review some basics of Carrollian amplitudes and compute several examples of Carrollian amplitudes for external massless scalar particles in three dimensions. We mainly focus on four-point amplitudes with contact and exchange diagrams and show the double copy structure between the Carrollian four-point amplitudes of Chern-Simons theory coupled to scalar matter and scalars minimally coupled to gravity. In section \ref{sec:3}, we review some basics of (planar) Bondi coordinates for flat space and AdS$_3$ and show how the bulk flat space limit is implemented as the Carrollian limit $c\rightarrow0$ at the boundary. We show how in the flat space limit, the bulk-to-bulk and bulk-to-boundary propagators of Witten diagrams reduce to those in Feynman diagrams. In section \ref{sec:4}, using the prescription in \cite{Alday:2024yyj}, we show in detail how to reproduce the Carrollian amplitudes that we obtain in Section \ref{sec:2} by taking the Carrollian limit of holographic correlators in AdS$_3$/CFT$_2$. We also discuss the two-point correlator in AdS$_3$ shockwave backgrounds. In section \ref{sec:5}, we discuss briefly some future directions.

\section{Carrollian amplitudes for massless scatterings} \label{sec:2}
In \cite{Donnay:2022wvx,Mason:2023mti}, it was shown that scattering amplitudes of massless particles in $d$-dimensional flat space can be recast as Carrollian correlators in $d-1$ dimensions by Fourier transforms. The Carrollian correlators living at null infinities are called Carrollian amplitudes. We shall focus on three-dimensional flat space that we will parametrize using planar Bondi coordinates. The planar Bondi coordinates for four-dimensional flat space were recently used in \cite{He:2019jjk,Gonzo:2020xza,Alday:2024yyj}. One can find the Bondi coordinates in three dimensions in \cite{Barnich:2012aw}. Here, we review some basics of them. The coordinates used are given by $u,r,z\in \mathbb{R}$. One can express the usual flat-space Cartesian coordinates in terms of the planar Bondi coordinates,
\begin{equation}
    \begin{split}
        \mathrm{X}^{\mu}&= \frac{1}{4}u \partial_z^2 q^{\mu} +\frac{1}{2}r q^{\mu} \\
        &= \left(\frac{1}{2}u+\frac{1}{2}r(1+z^2), \, rz, \,  -\frac{1}{2}u+\frac{1}{2}r(1-z^2) \right) \, ,
    \end{split}
\end{equation}
using the parametrization of a null vector
\begin{equation}
    q^\mu(z)=(1+z^2,2z,1-z^2) \, , \label{eq:qmu}
\end{equation}
in terms of a point $z\in S^1$. The metric in the planar Bondi coordinates is \cite{Barnich:2012aw} 
\begin{equation}
    ds^2_{Flat} = - du dr +r^2 dz^2 \, .
\end{equation}
In these conventions, the spacetime interval between two points is written as
\begin{equation}
    \xi_{12}^{Flat} = \mathrm{X}_{12}^{\mu}(\mathrm{X}_{12})_{\mu} = -u_{12} r_{12} + r_1 r_2 z_{12}^2 \, .
\end{equation}
The future null infinity corresponds to sending $r\rightarrow+\infty$, while the past null infinity corresponds to $r\rightarrow -\infty$. At null infinity $r\rightarrow\pm \infty$, the boundary metric becomes degenerate
\begin{equation}
    q_{ab} = 0 du^2+dz^2 \, ,
\end{equation}
which corresponds to the Carrollian structure at null infinity $\mathcal{I}\cong \mathbb{R}\times S^1$, where we will treat $z$ as the decompactified coordinate for the celestial circle.

The bulk Poincar\'e symmetries are generated by vector fields
\begin{equation}
    \eta=(\mathcal{T}+u \alpha) \partial_u +\mathcal{Y}\partial_z \, , \quad \alpha = \partial_z \mathcal{Y} \, ,
\end{equation}
where $\mathcal{T}(z) = 1, z, z^2$ are the three translations, and $\mathcal{Y}(z)=1,z,z^2$ are the three Lorentz transformations. At the boundary, these correspond to the global part of Carrollian symmetries.
A Carrollian primary operator $\Phi_\Delta(u,z)$ is defined as an operator that transforms as 
\begin{equation}
    \delta_{(\mathcal{T},\mathcal{Y})}\Phi_\Delta = \left[ (\mathcal{T}+u \partial_z \mathcal{Y})\partial_u + \mathcal{Y}\partial_z +\frac{\Delta}{2} \partial_z \mathcal{Y}\right]\Phi_\Delta \, ,
\end{equation}
$\Phi_\Delta(u,z)$ is a Carrollian primary of conformal dimension $\Delta$.

Since our starting point will be amplitudes of massless particles, we choose a convenient parametrization for generic null momenta:
\begin{equation}
    p_i^\mu = \epsilon_i \omega_i q_i^{\mu} \, ,
\end{equation} 
where $\epsilon_i=\pm 1$ corresponds to outgoing or incoming particle and $\omega_i$ is the light-cone energy of the associated momentum. Adapting the conventions in \cite{Lam:2017ofc}, the spacetime signature we use in this section is $(-,+,+)$. The Lorentzian invariant inner product between two momenta is written as
\begin{equation}
    p_i \cdot p_j = -2\epsilon_i\epsilon_j \omega_i\omega_j z^2_{ij} \, ,
\end{equation}
where $z_{ij}=z_i-z_j$. Under the Lorentz symmetry $SL(2,R)$,
\begin{equation}
    z\rightarrow \frac{a z+b}{c z+d} \,  \quad (ad-bc=1) \, ,
\end{equation}
the light-cone energy transforms as
\begin{equation}
    \omega\rightarrow (c z+d)^2 \omega \, .
\end{equation}

Starting from the scattering amplitudes $\mathcal{A}_n(\{\omega_1, z_1\}_{\epsilon_1},\dots,\{\omega_n, z_n\}_{\epsilon_n})$ in the momentum basis, the Carrollian amplitudes depending on the coordinates $u_i$ and $z_i$ at null infinity  are obtained by performing Fourier transforms with respect to
the light-cone energies 
\begin{equation}
\begin{split}
       & C_n(\{u_1, z_1\}_{\epsilon_1}, \dots,\{u_n, z_n\}_{\epsilon_n} ) \\
        =& \prod_{i=1}^n \left( \int_0^{+\infty} \frac{d\omega_i}{2\pi} e^{i\epsilon_i \omega_i u_i}\right)\mathcal{A}_n(\{\omega_1, z_1\}_{\epsilon_1},\dots,\{\omega_n, z_n\}_{\epsilon_n}) \, , \label{eq:nCarrollian}
\end{split}
\end{equation}
where we neglected the labels of helicity as we mainly focus the cases where all the external particles are scalar.

These Carrollian amplitudes can be re-interpreted as correlators in a Carrollian 2D CFT at
null infnity \cite{Donnay:2022wvx,Mason:2023mti}:
\begin{equation}
C_n(\{u_1, z_1\}_{\epsilon_1}, \dots,\{u_n, z_n\}_{\epsilon_n} ) = \langle \Phi_{\Delta_1, \epsilon_1}(u_1,z_1) \ldots \Phi_{\Delta_n, \epsilon_n}(u_n,z_n) \rangle \, ,
\label{holographic identification}
\end{equation} 
where $\Phi_{\Delta_i, \epsilon_i}(u_i,z_i)$ are conformal Carrollian primaries with conformal dimensions $\Delta_i=1$.

Generically, the $\partial_u-$ descendants of conformal Carrollian primaries are also primaries. Each $\partial_u-$ derivative increases the conformal dimension of the operator by $1$. These primaries will turn out to appear naturally when taking a flat space limit of AdS Witten diagrams, while also having better convergence properties. One can obtain correlators of $\partial_u-$ descendants simply by  taking $u$ derivatives of the correlator (\ref{holographic identification}). Adopting the notations of \cite{Mason:2023mti}, we have:
\begin{align}
&C_n^{m_1\dots m_n}(\{u_1, z_1\}_{\epsilon_1}, \dots,\{u_n, z_n\}_{\epsilon_n}) = \partial_{u_1}^{m_1}\dots \partial_{u_n}^{m_n}C_n(\{u_1, z_1\}_{\epsilon_1}, \dots,\{u_n, z_n\}_{\epsilon_n}) \nonumber\\
&= \prod_{i=1}^n \left( \int_0^{+\infty} \frac{d\omega_i}{2\pi} \, (i\epsilon_i \omega_i)^{m_i} e^{i\epsilon_i \omega_i u_i}\right)\mathcal{A}_n(\{\omega_1, z_1\}_{\epsilon_1},\dots,\{\omega_n, z_n\}_{\epsilon_n}) \, .
\end{align}
For later use, we introduce a shorthand notation for $C_n^{1\dots1}$:
\be
\widetilde{C}_n(\{u_1, z_1\}_{\epsilon_1}, \dots,\{u_n, z_n\}_{\epsilon_n}) = C_n^{1\dots1}(\{u_1, z_1\}_{\epsilon_1}, \dots,\{u_n, z_n\}_{\epsilon_n}) \, . \label{eq:tildeC_n}
\ee
\par In the next subsections we will introduce examples of four-point scalar amplitudes at tree level, which will be the main focus of this work. The two- and three-point cases are similar to those discussed in \cite{Alday:2024yyj}.

\subsection{Contact diagram}
The four-point celestial amplitudes involving massless scalars in three dimensions have been studied in \cite{Lam:2017ofc}. Here, we adapt a similar convention of kinematics variables. The scattering amplitudes $\mathcal{A}_4(\{\omega_1, z_1\}_{\epsilon_1},\dots,\{\omega_n, z_n\}_{\epsilon_4})$ contain momentum-conserving $\delta$ functions
\begin{equation}
    \mathcal{A}_4(\{\omega_1, z_1\}_{\epsilon_1},\dots,\{\omega_4, z_4\}_{\epsilon_4}) = \mathcal{T}(s,t) \delta^{(3)}\left(\sum_{i=1}^4\epsilon_i \omega_i q_i^{\mu}\right) \, ,
\end{equation}
where $s$, $t$ and $u$ are Mandelstam variables defined as $s=-(p_1+p_2)^2$, $t=-(p_1+p_3)^2$, $u=-(p_1+p_4)^2$. Using the parameterization shown in (\ref{eq:qmu}), the $\delta$ functions can be written as
\begin{multline}
    \delta^{(3)}\left(\sum_{i=1}^4 \epsilon_i\omega_i q^\mu(z_i)\right)=\frac{1}{4 |z_{23}z_{24}z_{34}|}\delta\left(\omega_2+\epsilon_1\epsilon_2\frac{z_{13}z_{14}}{z_{23}z_{24}}\omega_1\right)\delta\left(\omega_3-\epsilon_1\epsilon_3\frac{z_{12}z_{14}}{z_{23}z_{34}}\omega_1\right)\times\\
    \times \delta\left(\omega_4+\epsilon_1\epsilon_4\frac{z_{12}z_{13}}{z_{24}z_{34}}\omega_1\right) \, ,
\end{multline}
where $z_{ij}=z_i-z_j$. On the support of the $\delta$ functions, the Mandelstam variables are
\begin{equation}
    s=-4 \frac{z_{12}^2 z_{13}z_{14}}{z_{23}z_{24}}\omega_1^2 \,, \quad t = -\frac{1}{z}s \, , \quad u = \frac{1-z}{z}s \, ,  \label{eq:stu}
\end{equation}
where the conformal invariant cross ratio is 
\begin{equation}
    z= \frac{z_{12}z_{34}}{z_{13}z_{24}} \, .
\end{equation}
For a 2-to-2 scattering process, one needs to specify the incoming and outgoing configurations, which correspond to the following constraints on the cross ratio,
\begin{equation}
    \begin{split}
        &a) \,12 \rightleftharpoons 34\, , \quad z>1 \, ,\\
        &b) \, 13 \rightleftharpoons 24\, , \quad  0<z<1 \, ,\\
        &c) \, 14 \rightleftharpoons 23\, , \quad z<0  \, .
    \end{split}
\end{equation}
We shall focus on the case where particles $1$ and $2$ are incoming, while particles $3$ and $4$ are outgoing. The scattering angle $\theta$ is related to the cross ratio as $z^{-1}=\sin^2(\theta/2)<1$. The other scattering channels can be analyzed in a similar way, see e.g., \cite{Lam:2017ofc} for a discussion on the associated celestial amplitudes.

We begin with the simplest case which is the contact diagram of four massless scalars,
\begin{equation}
    \mathcal{T}(s,t)_{\text{contact}} = \kappa_4 \, ,
\end{equation}
where $\kappa_4$ is the coupling constant for a $\phi^4$ interaction. Using the definition in (\ref{eq:nCarrollian}), the Carrollian amplitude is:
\begin{equation}
    \begin{split}
        C_{4,\text{contact}} &= \frac{\kappa_4}{(2\pi)^4}\frac{1}{4 |z_{23}z_{24}z_{34}|}\int_0^\infty d\omega_1\exp{\left(-i \omega_1\left[u_1-u_2\frac{z_{13}z_{14}}{z_{23}z_{24}}+u_3\frac{z_{12}z_{14}}{z_{23}z_{34}}-u_4\frac{z_{12}z_{13}}{z_{24}z_{34}}\right]\right)} \\
        &=\frac{-i\kappa_4}{(2\pi)^4}\frac{\text{sign}(z_{23}z_{24}z_{34})}{4 z_{23}z_{24}z_{34}} \frac{1}{u_1-u_2\frac{z_{13}z_{14}}{z_{23}z_{24}}+u_3\frac{z_{12}z_{14}}{z_{23}z_{34}}-u_4\frac{z_{12}z_{13}}{z_{24}z_{34}}} \, . \label{eq:C4contact}
    \end{split}
\end{equation}
The four-point Carrollian amplitude (\ref{eq:C4contact}) is time-dependent, corresponding to the electric branch of Carrollian CFT. In the literature of Carrollian theories, there are two  distinct branches of solutions corresponding to two different types of Carroll invariant actions, which are referred to as the timelike (electric) and spacelike (magnetic) theories \cite{Duval:2014uoa,Henneaux:2021yzg,deBoer:2021jej,Hansen:2021fxi,Baiguera:2022lsw}. It has been shown that in (1+3) dimensions, the Carrollian amplitudes associated with scattering configurations belong to the electric branch \cite{Donnay:2022wvx,Mason:2023mti}. Our findings here confirm that this is also the case for (1+2) dimensions.

One can check that the Carrollian amplitude  (\ref{eq:C4contact}) satisfies the Ward identities generated by $M_n=z^{n+1}\partial_u$ with $n=-1,0,1$. These corresponds to the global part of the BMS$_3$ supertranslation symmetry. Compared to the four-point Carrollian amplitudes in \cite{Alday:2024yyj}, (\ref{eq:C4contact}) is non-distributional as the bulk translation symmetry is less constraining in three dimensions.

In Section \ref{sec:4}, we will show that the Carrollian amplitude (\ref{eq:C4contact}) can be reproduced by taking the Carrolllian limit $c\rightarrow0$ of the corresponding boundary correlator in AdS$_3$/CFT$_2$. We shall see that the analytic continuation to the Lorentzian signature is essential in this matching as shown in AdS$_4$/CFT$_3$ in \cite{Alday:2024yyj}. The $c\rightarrow0$ limit extracts the universal leading singularity upon the analytic continuation.

\subsection{Exchange diagrams}
In this section, we consider several examples of exchange diagrams. We begin with the simplest case of scalar exchange. For simplicity we only show the $s$-channel diagram
\begin{equation}
    \mathcal{T}_{s,\text{scalar}} =  \frac{\kappa_3^2}{s} = - \kappa_3^2 \frac{z_{23}z_{24}}{4 z_{12}^2 z_{13} z_{14} \omega_1^2} \, ,
\end{equation}
where we expressed the $s$ Mandelstam variable using (\ref{eq:stu}) and $\kappa_3$ is the coupling constant for a $\phi^3$ interaction. The Carrollian amplitude associated to scalar exchange is
\begin{equation}
\begin{split}
    C_{4,s, \text{scalar}}=&-\frac{\kappa_3^2}{(4\pi)^4}\frac{1}{|z_{23}z_{24}z_{34}|}\frac{z_{23}z_{24}}{z_{12}^2 z_{13} z_{14}} \\
    &\times\int_0^\infty \frac{d\omega_1}{\omega_1^2} \exp{\left(-i \omega_1\left[u_1-u_2\frac{z_{13}z_{14}}{z_{23}z_{24}}+u_3\frac{z_{12}z_{14}}{z_{23}z_{34}}-u_4\frac{z_{12}z_{13}}{z_{24}z_{34}}\right]\right)} \, ,
\end{split}
\end{equation}
which is IR divergent. One way to regularize it is to compute the $\partial_u$ descendants of the Carrollian amplitude. We are interested in 
\begin{equation}
\begin{split}
    &\widetilde{C}_{4,s, \text{scalar}}=\partial_{u_1}\partial_{u_2}\partial_{u_3}\partial_{u_4}C_{4,s, \text{scalar}} \\
    =& -\frac{\kappa_3^2}{(4\pi)^4}\frac{\text{sign}(z_{23}z_{24}z_{34})}{z_{12}^2 z_{13}z_{14}z_{34}} \int_0^{+\infty} d\omega_1 \omega_1^2 \left(\frac{z_{13}z_{14}}{z_{23}z_{24}}\right)\left(\frac{z_{12}z_{14}}{z_{23}z_{34}}\right) \left(\frac{z_{12}z_{13}}{z_{24}z_{34}} \right)\\
    &\times \exp{\left(-i \omega_1\left[u_1-u_2\frac{z_{13}z_{14}}{z_{23}z_{24}}+u_3\frac{z_{12}z_{14}}{z_{23}z_{34}}-u_4\frac{z_{12}z_{13}}{z_{24}z_{34}}\right]\right)} \\
    =& -\frac{\kappa_3^2 \text{sign}(z_{23}z_{24}z_{34})}{(4\pi)^4}\frac{z_{13}z_{14}}{z_{23}^2 z_{24}^2 z_{34}^3} \frac{2i}{\left(u_1-u_2\frac{z_{13}z_{14}}{z_{23}z_{24}}+u_3\frac{z_{12}z_{14}}{z_{23}z_{34}}-u_4\frac{z_{12}z_{13}}{z_{24}z_{34}} \right)^3} \, . \label{eq:tC4scalar}
\end{split}
\end{equation}
As we mentioned before, one can check that the conformal dimensions for each operator in $\widetilde{C}_4$ is $\Delta_i=2$. 
%In Section \ref{sec:3}, we will show that (\ref{eq:tC4scalar}) can be reproduced by the Carrollian limit of the corresponding Witten diagram of $s$ channel scalar exchange after analytic continuation to the Lorentzian signature.

Scalar exchange is not the only option that can be studied. One can look at 3D Chern-Simons (CS) theory with scalar matter, where the exchanged field will be a gauge boson. The Feynman rules in Lorenz gauge can be found in \cite{Ben-Shahar:2021zww}. The $s$ channel gauge field exchange diagram contribution is
\begin{equation}
\begin{split}
    \mathcal{T}_{s\text{,CS}}  &=g^2\frac{\epsilon^{\mu\nu\rho}(p_1-p_2)_\mu(p_3-p_4)_\nu(p_1+p_2)_\rho}{2\, (p_1+p_2)^2}= g^2\frac{\epsilon^{\mu\nu\rho}p_{1,\mu}p_{2,\nu}p_{4,\rho}}{\langle12\rangle^2}\\
    &=\frac{g^2}{2} \frac{\langle13\rangle\langle14\rangle}{\langle 34\rangle} = g^2 \omega_1\frac{z_{13}z_{14}}{z_{34}}\, , \label{eq:T4CS}
\end{split}
\end{equation}
where we used the following parametrization for the spinor-helicity variables in 3D $p_{\alpha\beta}=\lambda_{\alpha}\lambda_{\beta}$ with $\lambda_{\alpha}=\sqrt{2\omega}(z,1)$ and $g$ is the coupling constant.

The Carrollian amplitude is: 
\begin{equation}
 C_{4,s, \text{CS}}   =-\frac{g^2 \text{sign}(z_{23}z_{24}z_{34})}{(2\pi)^4}\frac{z_{13}z_{14}}{4 z_{23}z_{24}z_{34}^2}\frac{1}{\left(u_1-u_2\frac{z_{13}z_{14}}{z_{23}z_{24}}+u_3\frac{z_{12}z_{14}}{z_{23}z_{34}}-u_4\frac{z_{12}z_{13}}{z_{24}z_{34}}\right)^2}  \label{eq:C4CS}
\end{equation}

One can also look at the exchange of a graviton minimally coupled to scalars. In three dimensions there is no dynamical degree of freedom for gravity. However, one can still compute the amplitude of external massless scalars with a non-dynamical gravitational field being exchanged. One finds \cite{Gary:2009ae,Barker:1966zz,Huggins:1987ea}
\begin{equation}
    \mathcal{T}_{s,\text{GR}}= \kappa^2 \frac{t^2+st}{-s} = \kappa^2  \frac{\langle13\rangle^2\langle14\rangle^2}{\langle 34\rangle^2} \, , \label{eq:T4GR}
\end{equation}
where $\kappa$ is the coupling constant. Interestingly, (\ref{eq:T4GR}) is the double copy of (\ref{eq:T4CS}).
Using (\ref{eq:stu}), it becomes
\begin{equation}
    \mathcal{T}_{s,\text{GR}} = 4\kappa^2 \omega_1^2 \frac{z_{13}^2 z_{14}^2}{z_{34}^2} \, .
\end{equation}
Its Carrollian ampliltude is then:
\begin{equation}
    C_{4,s, \text{GR}} 
    =\frac{\kappa^2 \text{sign}(z_{23}z_{24}z_{34})}{(2\pi)^4} \frac{z_{13}^2 z_{14}^2}{z_{23}z_{24}z_{34}^3} \frac{2i}{\left(u_1-u_2\frac{z_{13}z_{14}}{z_{23}z_{24}}+u_3\frac{z_{12}z_{14}}{z_{23}z_{34}}-u_4\frac{z_{12}z_{13}}{z_{24}z_{34}}\right)^3} \,. \label{eq:C4GR}
\end{equation}
In Section \ref{sec:4}, we will show that the above exchange amplitudes,  (\ref{eq:C4contact}) (\ref{eq:tC4scalar}) (\ref{eq:C4CS}) and (\ref{eq:C4GR}), can be reproduced from the Carrollian/flat space limit of their corresponding holographic correlators in AdS$_3$/CFT$_2$. 

We also notice an interesting double copy relation among the gauge theory and gravity exchange Carrollian amplitudes. From (\ref{eq:C4contact}), (\ref{eq:C4CS}) and (\ref{eq:C4GR}), we find
\begin{equation}
    C_{4,s, \text{GR}} =C_{4,\text{contact}}^{-1}  \, (C_{4,s, \text{CS}} )^2 \, ,
\end{equation}
provided the following identification of the coupling constants,
\begin{equation}
    \kappa^2= \frac{g^4}{8\kappa_4} \, .
\end{equation}
The inverse of $C_{4,\text{contact}}$ serves as the KLT kernel.

\section{Flat limit of propagators of Witten diagrams} \label{sec:3}
In this section, we introduce the Bondi coordinates in AdS$_3$ and see how their flat space limit lands on the planar Bondi coordinates for $\mathbb{R}^{1,2}$. Geometrically, in these coordinates, the flat limit on AdS$_3$ corresponds to the Carrollian limit of the boundary. The building blocks of dynamical statements are the propagators appearing in Feynman or Witten diagrams. We shall see that after performing the flat space limit in Bondi coordinates, there is a natural identification between the two types of propagators.

\subsection{Planar Bondi Coordinates}

The planar Bondi coordinates parametrizing flat space also exist in AdS. One can see them by expressing the AdS metric in embedding coordinates. For AdS$_3$, they parametrize a hyperboloid in four dimensional flat space $X^I=(X^+, X^-, X^0, X^1)$ with the metric
\begin{equation}
    G_{IJ}dX^IdX^J= -dX^+dX^- - (dX^0)^2 +(dX^1)^2 \, . \label{eq:GIJ}
\end{equation}
AdS$_3$ is the hyperboloid defined by 
\begin{equation}
    X\cdot X = -X^+ X^- - (X^0)^2+ (X^1)^2 = -l^2 \, .
\end{equation}
In this section, we focus on {\it Lorentzian} AdS$_3$, where there are two time-like directions in the embedding coordinates. The planar Bondi coordinates for AdS$_3$ are obtained as the specific parametrization:
\begin{equation}
    X^I = r\left(1, \frac{u}{r}-\frac{u^2}{4l^2}+z^2, -\frac{l}{r}+\frac{u}{2l}, z\right) \, . \label{eq:embeddingXI}
\end{equation}
One can check that the parametrization satisfies the definition for AdS$_3$, $G_{IJ}X^IX^J=-l^2$ with $l$ being the AdS radius.

Using (\ref{eq:embeddingXI}), one finds the metric in the planar Bondi coordinates,
\begin{equation}
    ds^2_{AdS3} = -\frac{r^2}{4l^2}du^2 - du dr+ r^2 dz^2 \, . 
\end{equation}
The conformal boundary is reached as $r\rightarrow \pm \infty$ with metric
\begin{equation}
    ds^2_{\partial AdS3} = -\frac{1}{4l^2} du^2 + dz^2 \, .
\end{equation}
Both the bulk and boundary metrics reduce to those in the flat case as $l\rightarrow\infty$. In particular, one can identify the speed of light on the boundary $c$ as $1/l$. The boundary metric becomes degenerate as $l\rightarrow\infty$, which corresponds to the Carrollian limit $c\rightarrow0$ limit on the boundary.

The chordal distance defined by the embedding coordinates can be expressed in terms of the planar Bondi coordinates:
\begin{equation}
    \xi^{AdS}_{12} = (X_1-X_2)\cdot(X_1-X_2) = -\frac{r_1r_2}{4l^2}u_{12}^2 - u_{12} r_{12} +r_1 r_2 z_{12}^2 \, .
\end{equation}
In the flat space limit $l\rightarrow\infty$ it reduces to the desired invariant distance $\xi^{AdS}_{12} \rightarrow \xi^{Flat}_{12}$.

We can also consider Euclidean AdS$_3$, which amounts to changing the sign in front of $(dX^0)^2$ in (\ref{eq:GIJ}).

%The embedding coordinates (43) have two time like directions. In the flat space limit, one of the time like coordinates has to decouple and we are left with three coordinates. In (43) in the large $l$ limit, $X^0 \rightarrow \infty$, which is the one that decouples. What is left becomes the flat space coordinates (38).

\subsection{Propagators of Witten diagrams}
The basic building blocks of boundary correlators in AdS are bulk-to-bulk and bulk-to-boundary propagators. Similarly, the building blocks of flat space amplitudes are Feynman propagators. One crucial first step in identifying the flat limit of AdS boundary correlators is matching the flat limit of bulk-to-bulk or bulk-to-boundary to Feynman propagators in position space. In this section we will show that this matching holds in AdS$_3$.\\

We start from the scalar bulk-to-bulk propagator which is a solution to the equation of motion:
\begin{equation}
    (\Box_1+m^2)G_{BB}^{AdS,\Delta}(X_1,X_2)=\frac{1}{\sqrt{-g}}\delta^{(4)}(X_{12}),
\end{equation}
where $m^2l^2=\Delta(\Delta-2)$ and the d'Alembertian in Bondi coordinates is:
\begin{equation}
    \Box=\left[\frac{1}{r^2}\partial_z^2-4\partial_u\partial_r-\frac{2}{r}\partial_u+\frac{r}{l^2}(3\partial_r+r\partial_r^2)\right].
\end{equation}
The bulk-to-bulk propagator will be a function of the chordal distance $\xi_{12}^{AdS}$ and we perform a change of variables to the dimensionless variable:
\begin{equation}
    \chi_{12}=-\frac{4l^2}{\xi_{12}^{AdS}}.
\end{equation}
The solution to the wave equation then reads
\cite{Penedones:2010ue, DHoker:1999mqo}:
\begin{equation}
  G_{BB}^{AdS,\Delta}(X_1,X_2) =   C(\Delta)(\chi_{12})^\Delta\ _2F_1(\Delta,\Delta-\frac{1}{2},2\Delta-1;\chi_{12}) \, , \label{eq:GBBAdS}
\end{equation}
with 
\begin{equation}
    C(\Delta)=-\frac{i(-1)^\Delta }{2\pi\times 4^\Delta l},
\end{equation}
where the $-i$ difference compared to \cite{Penedones:2010ue, DHoker:1999mqo} is due to Wick rotation from EAdS$_3$ and a different convention for the Green's function.

We are interested now in the flat space limit. This is obtained by taking $l\rightarrow\infty$ and thus $\chi_{12}\rightarrow\infty$. Note that for our purposes, $\Delta$ does not rescale with $l$ as we take the flat space limit. Therefore, all scalars become massless regardless of their conformal dimension. The dominant term in the expansion of the hypergeometric function near infinity is given by:
\begin{equation}
    \frac{1}{(\chi_{12})^{\Delta-\frac{1}{2}}}\frac{i(-1)^\Delta \sqrt{\pi}\Gamma(2\Delta-1)}{\Gamma(\Delta)\Gamma(\Delta-\frac{1}{2})} \, .
\end{equation}
So in the $\chi_{12}\rightarrow \infty$ limit, the propagator (\ref{eq:GBBAdS}) becomes:
\begin{equation}
\begin{split}
  G_{BB}^{AdS,\Delta}(X_1,X_2) \xrightarrow{\chi_{12}\rightarrow\infty} &\frac{1}{8\pi}\frac{\chi_{12}^{\frac{1}{2}}}{l}=\frac{1}{4\pi}\frac{1}{\sqrt{-\xi_{12}^{AdS}-i \epsilon}}\\
  =&\frac{1}{4\pi}\frac{1}{\sqrt{\frac{r_1r_2}{4l^2}u_{12}^2 + u_{12} r_{12} -r_1 r_2 z_{12}^2-i\epsilon}} \, ,
\end{split}
\end{equation}
where the analytic continuation to the Lorentzian signature is implemented by $\xi_{12}^{AdS}\rightarrow\xi_{12}^{AdS}+i\epsilon \, $(see e.g. \cite{Duffin}). Taking $l\rightarrow\infty$ in the last line leads to
\begin{equation}
  G_{BB}^{AdS,\Delta}(X_1,X_2)   \xrightarrow{l\rightarrow\infty, \text{Bondi}}\frac{1}{4\pi}\frac{1}{\sqrt{u_{12} r_{12}-r_1 r_2 z_{12}^2-i\epsilon}} \, ,
\end{equation}
which agrees with the flat space Feynman propagator for massless scalar in position space \cite{Zhang:2008jy}. 

We conclude that in Bondi coordinates, the AdS$_3$ massive scalar bulk-to-bulk propagator reduces to the massless Feynman propagator in the limit $l\rightarrow\infty$. In particular, the parameter $\Delta$ completely disappears. This property is the same as that in \cite{Alday:2024yyj} for the AdS$_4$ case.

The bulk-to-boundary propagator in AdS can be obtained from the bulk-to-bulk propagator by sending one of the bulk points to the boundary. This means that $\xi_{12}^{AdS}\rightarrow\infty$ and $\chi_{12}\rightarrow 0$. Focusing on the first field being sent on the boundary, we see that the propagator (\ref{eq:GBBAdS}) becomes:
\begin{equation}
   G_{BB}^{AdS,\Delta}(X_1,X_2) \xrightarrow{\chi_{12}\rightarrow 0}  C(\Delta)(\chi_{12})^\Delta=C(\Delta)\left(\frac{l}{r_1}\right)^\Delta\left(\frac{-4l}{-\frac{r_2}{4l^2}u_{12}^2-u_{12}+r_2z_{12}^2+i\epsilon}\right)^\Delta \, .
\end{equation}
As we discussed in this section, the future null infinity (outgoing) corresponds to sending $r_1\rightarrow+\infty$, while the past null infinity (incoming) corresponds to $r_1\rightarrow -\infty$. The bulk-to-boundary propagator is taken for the $r_1\rightarrow+\infty$ limit of the above:
\begin{equation}
  G_{\partial B}^{AdS,\Delta}(x,X) =\lim\limits_{r_1\rightarrow\infty} l^{1/2}\left(\frac{r_1}{l}\right)^\Delta G_{BB}^{AdS,\Delta}(X_1,X_2)= l^{1/2}C(\Delta)\left(\frac{-4l}{-\frac{r_2}{4l^2}u_{xX}^2-u_{x}-q_x\cdot X +i\epsilon}\right)^\Delta \, .
\end{equation}
In the flat limit, this reduces to:
\begin{equation}
  G_{\partial B}^{AdS,\Delta} (x,X)\xrightarrow{l\rightarrow\infty, \text{Bondi}} \frac{-\alpha(\Delta)}{2\pi}\frac{\Gamma(\Delta)}{(-u-q\cdot X+i\epsilon)^\Delta} \, ,
\end{equation}
with
\begin{equation}
    \alpha(\Delta)=-(-1)^\Delta 4^\Delta 2\pi \frac{C(\Delta)}{\Gamma(\Delta)}l^{\Delta+\frac{1}{2}}=\frac{il^{\Delta-1/2}}{\Gamma(\Delta)} \, .
\end{equation}
For the case $\Delta=1/2$, the expression becomes the flat space Feynman propagator for massless scalars in position space \cite{Zhang:2008jy}, where one of the operators is sent to null infinity. In Sections \ref{sec:2} and \ref{sec:4}, we will mostly consider the case where $\Delta=1$ for simplicity. The other integer values of $\Delta$ do not provide any new data on flat space as they can be obtained by $\partial_u$ derivatives on the Carrollian primaries. This feature can also be found in \cite{Alday:2024yyj} for the AdS4
case. The bulk-to-boundary propagator of incoming particles can be obtained in a similar way. 

We have seen how the building blocks of Witten diagrams reduce to those of Feynman diagrams for the case of scalars. The spinning propagators can be found in \cite{DHoker:1999bve}. One can perform a similar analysis to that we have in this section and reach the same conclusion.

In the next section, we show how the Carrollian limit can be taken at the level of the holographic boundary correlators and its relation to the bulk point singularity.

\section{AdS$_3$ Witten diagrams, bulk point singularity and Carrollian limit} \label{sec:4}

In the previous section, we saw how the various ingredients contributing to Witten diagrams from a bulk perspective behave under the flat space limit. From a boundary perspective, in \cite{Alday:2024yyj}, it was shown that the Carrollian limit of the AdS$_4$ boundary CFT correlators reproduces the associated Carrollian amplitudes, which is equivalent to saying that the boundary Carrollian limit corresponds to the bulk flat limit. From a bulk perspective, the flat space amplitude can be recovered from the residue of the bulk-point singularity signaling the locality of the bulk interactions. To see the appearance of the bulk-point singularity on the boundary correlators, a key step in the derivation is performing an analytic continuation from the Euclidean signature of the usual Witten diagrams to the Lorentzian signature. The boundary Carrollian limit of the Lorentzian correlators focuses on and extracts the bulk point singularity contribution. In this section, we apply the same prescription to 2D holographic correlators in AdS$_3$/CFT$_2$, and show how to reproduce the four-point Carrollian amplitudes obtained in Section \ref{sec:2}. As we will see, there is no restriction on the boundary operator insertions a priori.

In the section, Euclidean correlatos will be denoted by $\langle \dots \rangle_E$. Lorentzian correlators will not have any subscripts. We will use $x=(ct, z)$ as Euclidean coordinates with metric $ds_E^2=c^2 dt^2+ dz^2$. The Euclidean time will be analytically continued to the Lorentzian time $u$.

\subsection{Contact diagram}
We start by considering a $\phi^4$ interaction in AdS$_3$ with coupling $\kappa_4$, the four-point boundary correlators are given by 
\begin{equation}
\begin{split}
   & \langle O_{\Delta_1}(x_1)O_{\Delta_2}(x_2)O_{\Delta_3}(x_3)O_{\Delta_4}(x_4)\rangle_E^{\text{contact}}= \kappa_4 \int_{AdS_3} d^3 X \prod_{i=1}^4 G_{\partial B}^{\Delta_i}(x_i,X) \, . \\
   &
\end{split}
\end{equation}
The integral expression defines $D$-functions (see e.g. \cite{Bissi:2022mrs} for a review on Witten diagrams),
\begin{equation}
\begin{split}
    &\langle O_{\Delta_1}(x_1)O_{\Delta_2}(x_2)O_{\Delta_3}(x_3)O_{\Delta_4}(x_4)\rangle_E^{\text{contact}} =\kappa_4 D_{\Delta_1,\Delta_2,\Delta_3,\Delta_4}(x_i)\\
    =&\kappa_4 \mathcal{N}_4 \frac{(x_{14}^2)^{\frac{1}{2}\Sigma_\Delta-\Delta_1-\Delta_4}(x_{34}^2)^{\frac{1}{2}\Sigma_\Delta-\Delta_3-\Delta_4}}{(x_{13}^2)^{\frac{1}{2}\Sigma_\Delta -\Delta_4} (x_{24}^2)^{\Delta_2}} \bar{D}_{\Delta_1,\Delta_2,\Delta_3,\Delta_4}(U,V) \, ,
\end{split}
\end{equation}
where 
\begin{equation}
    \mathcal{N}_4= \frac{\pi}{2} \Gamma\left( \frac{1}{2}\Sigma_\Delta-1\right)\prod_{i=1}^4 \frac{1}{\Gamma(\Delta_i)} \, ,\qquad \Sigma_\Delta= \sum_{i=1}^4 \Delta_i \, , \, 
\end{equation}
\begin{equation}
    U=\frac{x_{12}^2x_{34}^2}{x_{13}^2x_{24}^2},\ V=\frac{x_{23}^2x_{14}^2}{x_{13}^2x_{24}^2} \, ,
\end{equation}
where the normalization factors for bulk–to–boundary propagators are omitted \cite{Bissi:2022mrs}.
%\bnote {will double check the overall constant one we finish section 2 for the normalizations.}
%$\Delta_i$ is the conformal dimension of the CFT operator. Via AdS/CFT, it is related to the mass of the particle in the bulk by $m^2 = \Delta(\Delta-2)/l^2$. In the flat space limit $l \rightarrow\infty$, for a finite dimension $\Delta$, the associated particle becomes massless. This is the main focus of this work.

We will consider the simplest, $\Delta_i=1$ case. The $D$-function can be expressed in terms of elementary functions \cite{Denner:1991qq,Usyukina:1992jd}
\begin{equation}
    \langle O_{1}(x_1)O_{1}(x_2)O_{1}(x_3)O_{1}(x_4)\rangle_E^{\text{contact}} = \frac{\kappa_4 \pi}{2} \frac{1}{x_{13}^2x_{24}^2}\Bar{D}_{1,1,1,1}(U,V)  \, , \label{eq:Adscontact}
\end{equation}
with
\begin{equation}
    \Bar{D}_{1,1,1,1}(U,V)=\frac{1}{Z-\Bar{Z}}\left[2 \text{Li}_2(Z)-2\text{Li}_2(\Bar{Z})+\log(Z\Bar{Z})\log\left(\frac{1-Z}{1-\Bar{Z}}\right)\right] \, .
\end{equation}
We alternatively expressed the $U,V$ variables in terms of the usual boundary cross ratios
\begin{equation}
    U=Z\Bar{Z} \, , \quad V=(1-Z)(1-\Bar{Z}) \, .
\end{equation}

As written, the CFT correlator does not display any bulk-point singularity as it has been derived in an Euclidean regime. In order to connect to have a well-behaved flat space limit, we will need to analytically continue to the Lorentzian regime. In the Euclidean signature, the cross ratios $Z$ and $\bar{Z}$ are complex conjugated to each other. In the Lorentzian signature, they become independent. To perform the analytic continuation from Euclidean to Lorentzian, one starts from an Euclidean correlator and, depending on
the time ordering between the four operators in the Lorentzian spacetime, one chooses a path for the continuation. Recall that the scattering configuration considered in Section \ref{sec:2} is $12\rightarrow34$, which means that we shall place operators $3$ and $4$ in the future of operators $1$ and $2$. This corresponds to the following analytic continuation \cite{Lam:2017ofc},
\begin{equation}
    \text{12}\rightarrow \text{34}: \quad Z-1\rightarrow  e^{2\pi i}(Z-1), \quad \frac{1}{\bar{Z}} \rightarrow e^{2\pi i} \frac{1}{\bar{Z}} \, . \label{eq:EtoL}
\end{equation}
Monodromies of the polylogarithms around all branch points are well known (see e.g. \cite{Bourjaily:2020wvq,Alday:2024yyj}). One finds
\begin{equation}
    \log(1-Z) \rightarrow \log(1-Z)+2\pi i \, , \text{Li}_2(Z)\rightarrow \text{Li}_2(Z) -2\pi i \log(Z) \, ,
\end{equation}
\begin{equation}
\begin{split}
    \log(\bar{Z}) \rightarrow \log(\bar{Z})-2\pi i \, , \quad \log(1-\bar{Z}) \rightarrow \log(1-\bar{Z}) -2\pi i \, , \\
    \text{Li}_2(\bar{Z}) \rightarrow \text{Li}_2(\bar{Z})+2\pi i \log(\bar{Z}) +2\pi^2 \, .
\end{split}
\end{equation}
Therefore, after the analytic continuation (\ref{eq:EtoL}) the elementary function $\bar{D}_{1,1,1,1}$ receives an additional contribution 
\begin{equation}
    \bar{D}_{1,1,1,1}\rightarrow \bar{D}_{1,1,1,1} +\frac{4\pi^2}{Z-\bar{Z}}+\frac{2\pi i}{Z-\bar{Z}}\log\left( \frac{1-\bar{Z}}{1-Z}\right) \, .
\end{equation}
The important features of the new terms is that they are singular in the $Z\rightarrow\bar{Z}$ limit. This is relevant as we will see that on a kinematic basis, taking the boundary Carrollian limit is equivalent to the same $Z\rightarrow\bar{Z}$ limit. Without the singular terms, the Carrollian limit would be vanishing. We observe that the second term is subleading in the $Z\rightarrow\bar{Z}$ limit, but the first generates the leading singularity of the contact diagram:
\begin{equation}
    \hat{\Phi}_{1111}^{ls}  = \frac{4\pi^2}{Z-\bar{Z}} \, .
\end{equation}
The leading singularity corresponds to the bulk point singularity discussed in \cite{Gary:2009ae,Maldacena:2015iua}. Notice that depending on the path of analytic continuation, the final result might be different, see e.g. \cite{Alday:2024yyj} and \cite{Gary:2009ae}. However, when the leading singularity exists, it is independent of the path.

To connect Witten diagrams to flat-space Carrollian correlators, we note that in the Carrollian limit:
\begin{equation}
\begin{split}
    &(Z-\Bar{Z})^2=((1+U-V)^2-4U)\\
    \xrightarrow{c\rightarrow 0}&\frac{4c^2}{z_{13}^4 z_{24}^4} [u_4 z_{12}z_{13}z_{23}-u_3 z_{12}z_{14}z_{24}+u_2 z_{13}z_{14}z_{34}-u_1 z_{23}z_{24}z_{34}]^2+\mathcal{O}(c^4) , \label{eq:ZmZbar}
\end{split}
\end{equation}
where we used $x_{ij}^2=-c^2(u_i-u_j)^2+(z_i-z_j)^2$.
In contrast to the Carrollian limit of the cross-ratio difference in four dimensions \cite{Alday:2024yyj}, there is no constant term as $c\rightarrow0$ and the $Z-\Bar{Z}$ difference vanishes.
Taking the Carrollian limit $c\rightarrow0$ on the boundary leads to the limit $Z\rightarrow\bar{Z}$ in (\ref{eq:ZmZbar}). As $c\rightarrow0$, the contributing term will thus be the leading singularity discussed above. We then find that after analytic continuation to the Lorentzian signature, in the Carrollian limit $c\rightarrow 0$, the contact Witten diagram (\ref{eq:Adscontact}) becomes
\begin{equation}
\begin{split}
    &\langle O_{1}(x_1)O_{1}(x_2)O_{1}(x_3)O_{1}(x_4)\rangle^{\text{contact}} \\
    =&\kappa_4  \frac{\pi^3}{c |z_{23}z_{24}z_{34}|} \frac{1}{u_1-u_2\frac{z_{13}z_{14}}{z_{23}z_{24}}+u_3\frac{z_{12}z_{14}}{z_{23}z_{34}}-u_4\frac{z_{12}z_{13}}{z_{24}z_{34}}} \, ,
\end{split}
\end{equation}
which matches the Carrollian amplitude of the contact diagram in (\ref{eq:C4contact}) up to a constant factor.

A comment is in order regarding the choice of the conformal dimensions of the primary operators. It was shown that from the irreducible representations of the Poincar\'e group at null infinity, the natural scaling dimension would be $\Delta=\frac{1}{2}$ for AdS$_3$ in the flat limit \cite{Nguyen:2023vfz}. One can compute the Carrollian amplitude with $\Delta=\frac{1}{2}$ with its $\partial_u$ descendants and match it to the corresponding Carrollian limit of the AdS$_3$ boundary correlator. The main difficulty is that there is no known expression for the corresponding AdS$_3$ boundary correlator in this case. However, the leading singularity of interest might be extracted following Appendix B in \cite{Alday:2024yyj}. We leave a detailed analysis for future work.

Next, we shall follow the same prescription to compute several examples of the Carrollian limit of boundary correlators in AdS$_3$/CFT$_2$ with integer dimensions and we will always see  the matching with the associated Carrollian amplitudes.

\subsection{Exchange diagrams}

The four-point correlator associated with a scalar exchange in the bulk is
\begin{equation}
    \langle O_{\Delta_1}(x_1)O_{\Delta_2}(x_2)O_{\Delta_3}(x_3)O_{\Delta_4}(x_4)\rangle_E^{\text{scalar exchange}}= I_s+I_t+I_u \, ,
\end{equation}
where $I_s$ corresponds to the $s$-channel exchange diagram with the exchange operator having conformal dimension $\Delta$,
\begin{equation}
    I_s = \kappa_3^2 \int_{AdS_3} d^3X d^3Y \prod_{i=1}^2 G^{\Delta_i}_{\partial B}(x_i,X)\prod_{i=3}^4 G_{\partial B}^{\Delta_i}(x_i,Y) \,  G_{BB}^\Delta(X,Y) \, ,
\end{equation}
and $\kappa_3$ is the coupling for the $\phi^3$ interaction. $I_t$ and $I_u$ can be obtained by the replacements $1\leftrightarrow3$ and $1\leftrightarrow4$ respectively. Exchange diagrams can not be written as elementary functions for generic conformal dimensions. However, when $\Delta_3+\Delta_4-\Delta\in 2 \mathbb{Z}^+$, they can be written as a finite sum of contact diagrams,
\begin{equation}
\begin{split}
     I_s &= \kappa_3^2 \sum_{k=k_\text{min}}^{k_{\text{max}}} a_k (x_{34}^2)^{k-\Delta_4} D_{\Delta_1,\Delta_2,\Delta_3-\Delta_4+k,k} \\
     &= \frac{(x_{14}^2)^{\frac{1}{2}\Sigma_\Delta-\Delta_1-\Delta_4}(x_{34}^2)^{\frac{1}{2}\Sigma_\Delta-\Delta_3-\Delta_4}}{(x_{13}^2)^{\frac{1}{2}\Sigma_\Delta -\Delta_4} (x_{24}^2)^{\Delta_2}}  \sum_{k=k_\text{min}}^{k_{\text{max}}} a_k \mathcal{N}'_4 \bar{D}_{\Delta_1,\Delta_2,\Delta_3-\Delta_4+k,k} \, ,
\end{split}
\end{equation}
where $k_{\text{min}} = \frac{1}{2}(\Delta-\Delta_3+\Delta_4)$, $k_{\text{min}} = \Delta_4-1$. The coefficients $a_k$ can be recursively. See \cite{Bissi:2022mrs} for explicit expressions for $a_k$ and $\mathcal{N}'_4$.

In order to match the Carrollian correlator (\ref{eq:tC4scalar}), we are interested in the case $\Delta_1=\Delta_2=\Delta_3=\Delta_4=\Delta=2$ for which we have
\begin{equation}
    I_{s}= \frac{\kappa_3^2}{(x_{13}^2)^2 (x_{24}^2)^2}\bar{D}_{2,2,1,1} \, ,
\end{equation}
where we ignored an overall normalization constant involving $\pi$.
Following the same analytic continuation as we had in (\ref{eq:EtoL}), the exchange diagram $\bar{D}_{2211}$ develops a leading singularity
\begin{equation}
    \bar{D}_{2,2,1,1} \rightarrow \bar{D}_{2,2,1,1}+ \hat{\Phi}^{ls}_{2211} +\cdots \, , \label{eq:Db2211}
\end{equation}
where the leading singularity $\hat{\Phi}^{ls}_{2211}$ can be obtained from $\hat{\Phi}^{ls}_{1111}$ by taking a derivative \cite{Alday:2024yyj},
\begin{equation}
    \hat{\Phi}^{ls}_{2211}=- \frac{2(1-Z)}{Z-\bar{Z}}\partial_{\bar{Z}}\hat{\Phi}^{ls}_{1111} = -8\pi^2\frac{(1-Z)}{(Z-\bar{Z})^3} \, . \label{eq:eq:ls2211}
\end{equation}
In the Carrollian limit $c\rightarrow0$, we have $(1-Z)\rightarrow \frac{z_{23}z_{14}}{z_{13}z_{24}}$, and $Z-\bar{Z}$ is given by (\ref{eq:ZmZbar}). Hence, the resulting correlator is
\begin{equation}
\begin{split}
    \lim_{c\rightarrow0} I_s &\sim\frac{\kappa_3^2}{z_{13}^4 z_{24}^4} \frac{z_{23}z_{14}}{z_{13}z_{24}} \frac{z_{13}^6 z_{24}^6}{c^3(z_{23}z_{24}z_{34})^3 \left(u_1-u_2\frac{z_{13}z_{14}}{z_{23}z_{24}}+u_3\frac{z_{12}z_{14}}{z_{23}z_{34}}-u_4\frac{z_{12}z_{13}}{z_{24}z_{34}} \right)^3} \\
    &=\frac{\kappa_3^2}{c^3} \frac{z_{13}z_{14}}{z_{23}^2 z_{24}^2 z_{34}^3} \frac{1}{\left(u_1-u_2\frac{z_{13}z_{14}}{z_{23}z_{24}}+u_3\frac{z_{12}z_{14}}{z_{23}z_{34}}-u_4\frac{z_{12}z_{13}}{z_{24}z_{34}} \right)^3}  \, ,
\end{split}
\end{equation}
which matches the associated Carrollian amplitude (\ref{eq:tC4scalar})  up to an overall constant. For the $t$ channel and $u$ channel exchange diagram, one can simply perform permutations.
\medskip

We consider next Chern-Simons theory with scalar matter in AdS$_3$. The massless Chern-Simons exchange diagram was computed in \cite{Rastelli:2019gtj}. For $\Delta_i=1$ the $s$ channel contribution is
\begin{equation}
    \langle O_1(x_1)O_1(x_2)O_1(x_3)O_1(x_4)\rangle_E^{\text{CS}} = \frac{1}{x_{12}^2 x_{34}^2} \left((Z-\bar{Z})U\bar{D}_{2,2,1,1} +\log(V) \right) \, . \label{eq:Wcs}
\end{equation}

Upon analytic continuation to the Lorentzian signature, the leading singularity comes from $\bar{D}_{2,2,1,1}$ that can be found in (\ref{eq:Db2211}) and (\ref{eq:eq:ls2211}).
The Carrollian limit of this exchange diagram (\ref{eq:Wcs}) gives us
\begin{equation}
    I_{s,\text{CS}} \sim \frac{1}{c^2}\frac{z_{13}z_{14}}{z_{23}z_{24}z_{34}^2}\frac{1}{\left(u_1-u_2\frac{z_{13}z_{14}}{z_{23}z_{24}}+u_3\frac{z_{12}z_{14}}{z_{23}z_{34}}-u_4\frac{z_{12}z_{13}}{z_{24}z_{34}}\right)^2}  \, ,
\end{equation}
which agrees with the corresponding Carrollian amplitude (\ref{eq:C4CS}).

Finally, the exchange Witten diagram of gravitons in AdS$_3$ can be found in \cite{Rastelli:2019gtj}. For $\Delta_i=1$ in the $s$ channel,
\begin{equation}
    \langle O_1(x_1)O_1(x_2)O_1(x_3)O_1(x_4)\rangle_E^{\text{GR}} = \frac{1}{ x_{12}^2x_{34}^2}(2+U(U-V-1)\bar{D}_{2,2,1,1}) \, 
\end{equation}
Once again the leading singularity comes from $\bar{D}_{2,2,1,1}$. We obtain the following Carrollian limit of the Lorentzian correlator,
\begin{equation}
   I_{s,\text{GR}} \sim \frac{1}{c^3}\frac{z_{13}^2 z_{14}^2}{z_{23}z_{24}z_{34}^3} \frac{1}{\left(u_1-u_2\frac{z_{13}z_{14}}{z_{23}z_{24}}+u_3\frac{z_{12}z_{14}}{z_{23}z_{34}}-u_4\frac{z_{12}z_{13}}{z_{24}z_{34}}\right)^3} \, ,
\end{equation}
which matches Carrollian amplitude (\ref{eq:C4GR}).

The same procedure can be repeated for higher conformal dimension $\Delta_i$ and the results will match the $\partial_u$-descendants of the associated Carrollian amplitude.

\subsection{Two-point correlators in AdS shockwave backgrounds}
In the previous sections we focused on tree level correlators on pure AdS$_3$ backgrounds. However, it is known that for certain kinematic regimes, quantum effects can be resummed to all orders in the relevant coupling constant. Initially studied in flat space, this is referred to as the eikonal regime, see, e.g. \cite{DiVecchia:2023frv} for a comprehensive review. One of the famous results shown by t'Hooft is that the 2-to-2 eikonal amplitude can be produced by the two-point amplitude in a shockwave background \cite{tHooft:1987vrq} where one of the particles can be viewed as the source of the shockwave. Similarly, in AdS, properties of the eikonal amplitude can be derived by studying the two-point function of a scalar in a shockwave background \cite{Cornalba:2006xk}. In this section we will analyze the behaviour of this background two-point scalar correlator under the Carrollian limit. The expression for the two-point correlator in the AdS$_{d+1}$ shockwave backgrounds has the following integral representation,
\begin{equation}
    \mathcal{E}=\int_0^\infty s^{2\Delta_1-1}ds\int_{H_{d-1}} \widetilde{dx} e^{2i q\cdot(sx)+(i s)^{j-1}h(x)} \, ,
\end{equation}
where we omit the overall normalization constant that can be found in \cite{Cornalba:2006xk}. $H_{d-1}$ is the transverse hyperbolic space determined by the shock, $h(x)$ is the metric deformation under the presence of the shock and it depends on Newton's constant $G$ and $j$ is spin of the exchange particle. See \cite{Cornalba:2006xk} for full details.

The expansion of the two-point correlator in Newton's constant $G$ takes the form
\begin{equation}
    \mathcal{E}= \mathcal{E}_0+\mathcal{E}_1+\cdots \, ,
\end{equation}
where $\mathcal{E}_1$ is the first non-trivial term that linearly depends on $G$.
Focusing on AdS$_3$, it was shown in \cite{Cornalba:2006xk}, for the case of $\Delta_1=\Delta_2=2$ and $j=0$, that one can evaluate all the integrals over $H_1$ to obtain a closed form expression for $\mathcal{E}_1$. The result is 
\begin{equation}
\begin{split}
    \mathcal{E}_{1}\sim M(w) &=- \frac{4G}{3} w \, _2F_1(1, 2, 4; 1-w) \\
    &= -\frac{4G\, w}{(1-w)^3}[1+2 w \ln(w) -w^2] \, , \label{eq:2ptshock}
\end{split}
\end{equation}
with $w$ defined in terms of the cross-ratios $Z$ and $\bar{Z}$: $w=Z/\bar{Z}$. We are interested in the Carrollian limit of the correlator. This is obtained by consider the bulk-point singular limit $Z\rightarrow \bar{Z}$. One finds that $Z\rightarrow \bar{Z}$ corresponds to $w\rightarrow 1$ in (\ref{eq:2ptshock}). In this limit, the two point is not singular
\begin{equation}
    \mathcal{E}_1 \sim 4G \, ,
\end{equation}
and would not match the Carrollian limit of the corresponding scalar exchange diagram which is order of $1/c^3$. 

This is consistent with the fact that in three-dimensional flat space there is no matching between four-point eikonal amplitudes and two-point amplitudes in a shockwave background. To see this, we remember that the regime of validity for the eikonal exponentiation is transparent in the impact parameter space,
\begin{equation}
    \frac{\hbar}{\sqrt{s}}\ll G \sqrt{s} b^{4-D} \ll b \, ,
\end{equation}
where $b$ is the impact parameter dual to the exchanged momentum $q$ and $D$ is the dimension of spacetime. It is straightforward to conclude that this condition cannot be satisfied in three dimensions $D=3$. The eikonal exponentiation exists only in four or higher dimensions.

Another non-trivial background where the flat limit prescription can be applied could be the BTZ black hole in three dimensions. Two-point functions of scalars have been computed in \cite{Keski-Vakkuri:1998gmz} and the flat limit of the BTZ geometry has been studied in \cite{Barnich:2012aw}. As BTZ spacetimes are quotients of AdS$_3$ by a Killing vector field, they are usually expressed in a convenient set of coordinates where the quotient procedure is trivialized. It would be interesting to understand the geometry and Carrollian limit in the planar Bondi coordinates and to further relate this to the bulk two-point function in the flat limit of BTZ.

\section{Concluding remarks} \label{sec:5}

There are many interesting and promising directions to explore in the near and far future. Here we list a few of them. 
It would be interesting to make a connection with the intrinsic Carrollian approach. See, e.g. \cite{deBoer:2023fnj,Chen:2024voz,Cotler:2024xhb,Kraus:2024gso,Kraus:2025wgi,Baiguera:2022lsw,Figueroa-OFarrill:2023qty,Ciambelli:2018wre,Ciambelli:2019lap,Cotler:2024cia}. In particular, \cite{Cotler:2024xhb} showed the existence of the infrared triangle in three dimensions. It would be interesting to show what the soft graviton theorem implies for Carrollian amplitudes of massless particles in three dimensions. To that end, one needs to consider spinning external particles. We hope to study it in the near future.

The four-point Carrollian amplitudes we computed are non-distributional, which would make it a perfect playground to test some Carrollian bootstrap ideas. It would be interesting to decompose our four-point Carrollian amplitudes into the BMS blocks shown in \cite{Bagchi:2016geg,Bagchi:2017cpu,Chen:2020vvn,Chen:2022cpx,Chen:2022jhx} and see how the geodesic Witten diagrams associated with the conformal blocks in \cite{Hijano:2015zsa,Hijano:2015qja} behave in the Carrollian picture. Another interesting direction is finding differential equations satisfied by Carrollian amplitudes in the same spirit as \cite{Ruzziconi:2024zkr}. We might find new solutions to the differential equations in comparison with the celestial version \cite{Fan:2022vbz}.

Perhaps the most exciting one, as mentioned in \cite{Alday:2024yyj}, is to see how far the prescription used in our work and \cite{Alday:2024yyj} can be pushed beyond holography. Carrollian correlators were computed efficiently without knowing the Lagrangian of the Carrollian theory, hence bypassing the difficulty in quantizing a Carrollian theory. It would be interesting to apply it to other Carrollian theories without a Lagrangian description. For instance, it would be fascinating to compute four-point and higher-point correlators in the Carrollian/BMS Ising model in e.g.\cite{Yu:2022bcp,Hao:2022xhq}.

\section*{Acknowledgements}
We would like to thank Tim Adamo, Romain Ruzziconi and Stephan Stieberger for valuable comments on the draft. We would like to thank Tim Adamo, Wei Bu, Eduardo Casali, Riccardo Gonzo, Vijay Nenmeli, Benjamin Strittmatter, Tom Taylor and Piotr Tourkine for useful discussions on related topics. IS is supported by an STFC studentship. BZ is supported by the Simons Collaboration on Celestial Holography.

%\appendix 
%\section{Other coordinates in AdS} \label{secAppA}

%-Global coordinates

%\noindent-Poincar\'e coordinates

%\section{Carrollian amplitudes in $(2,1)$ signature?}

%What is the analog of $(2,2)$ signature in 3D? is it $(2,1)$ signature?

%Even the Carrollian amplitudes in $(2,2)$ signature have not been worked out. At the level of Eq.(5.6) in \cite{Guevara:2021abz}.


\begin{thebibliography}{99}


%\cite{Alday:2024yyj}
\bibitem{Alday:2024yyj}
L.~F.~Alday, M.~Nocchi, R.~Ruzziconi and A.~Yelleshpur Srikant,
``Carrollian amplitudes from holographic correlators,''
JHEP \textbf{03}, 158 (2025)
doi:10.1007/JHEP03(2025)158
[arXiv:2406.19343 [hep-th]].
%27 citations counted in INSPIRE as of 24 Mar 2025


\bibitem{Raclariu:2021zjz} For recent reviews, see:\\
A.~M.~Raclariu,
``Lectures on Celestial Holography,''
[arXiv:2107.02075 [hep-th]].

%\cite{Pasterski:2021rjz}
%\bibitem{Pasterski:2021rjz}
S.~Pasterski,
``Lectures on celestial amplitudes,''
Eur. Phys. J. C \textbf{81}, no.12, 1062 (2021)
doi:10.1140/epjc/s10052-021-09846-7
[arXiv:2108.04801 [hep-th]].
%149 citations counted in INSPIRE as of 01 Dec 2023

%\cite{Strominger:2017zoo}
%\bibitem{Strominger:2017zoo}
A.~Strominger,
``Lectures on the Infrared Structure of Gravity and Gauge Theory,''
[arXiv:1703.05448 [hep-th]].
%723 citations counted in INSPIRE as of 14 Aug 2023



%\cite{Pasterski:2021raf}
%\bibitem{Pasterski:2021raf}
S.~Pasterski, M.~Pate and A.~M.~Raclariu,
``Celestial Holography,''
[arXiv:2111.11392 [hep-th]].
%104 citations counted in INSPIRE as of 03 Nov 2023

%\cite{Donnay:2023mrd}
%\bibitem{Donnay:2023mrd}
L.~Donnay,
``Celestial holography: An asymptotic symmetry perspective,''
Phys. Rept. \textbf{1073}, 1-41 (2024)
doi:10.1016/j.physrep.2024.04.003
[arXiv:2310.12922 [hep-th]].
%59 citations counted in INSPIRE as of 25 Mar 2025




\bibitem{Donnay:2022aba}
L.~Donnay, A.~Fiorucci, Y.~Herfray and R.~Ruzziconi,
``Carrollian Perspective on Celestial Holography,''
Phys. Rev. Lett. \textbf{129}, no.7, 071602 (2022)
doi:10.1103/PhysRevLett.129.071602
[arXiv:2202.04702 [hep-th]].
%130 citations counted in INSPIRE as of 11 Jun 2024

%\cite{Bagchi:2022emh}
\bibitem{Bagchi:2022emh}
A.~Bagchi, S.~Banerjee, R.~Basu and S.~Dutta,
``Scattering Amplitudes: Celestial and Carrollian,''
Phys. Rev. Lett. \textbf{128}, no.24, 241601 (2022)
doi:10.1103/PhysRevLett.128.241601
[arXiv:2202.08438 [hep-th]].
%91 citations counted in INSPIRE as of 11 Jun 2024

%\cite{Donnay:2022wvx}
\bibitem{Donnay:2022wvx}
L.~Donnay, A.~Fiorucci, Y.~Herfray and R.~Ruzziconi,
``Bridging Carrollian and celestial holography,''
Phys. Rev. D \textbf{107}, no.12, 126027 (2023)
doi:10.1103/PhysRevD.107.126027
[arXiv:2212.12553 [hep-th]].
%74 citations counted in INSPIRE as of 11 Jun 2024

%\cite{Banerjee:2018gce}
\bibitem{Banerjee:2018gce}
S.~Banerjee,
``Null Infinity and Unitary Representation of The Poincare Group,''
JHEP \textbf{01}, 205 (2019)
doi:10.1007/JHEP01(2019)205
[arXiv:1801.10171 [hep-th]].
%68 citations counted in INSPIRE as of 10 Jul 2024

%\cite{Banerjee:2019prz}
\bibitem{Banerjee:2019prz}
S.~Banerjee, S.~Ghosh, P.~Pandey and A.~P.~Saha,
``Modified celestial amplitude in Einstein gravity,''
JHEP \textbf{03}, 125 (2020)
doi:10.1007/JHEP03(2020)125
[arXiv:1909.03075 [hep-th]].
%32 citations counted in INSPIRE as of 10 Jul 2024




%\cite{Barnich:2012rz}
\bibitem{Barnich:2012rz}
G.~Barnich, A.~Gomberoff and H.~A.~Gonz\'alez,
``Three-dimensional Bondi-Metzner-Sachs invariant two-dimensional field theories as the flat limit of Liouville theory,''
Phys. Rev. D \textbf{87} (2013) no.12, 124032
doi:10.1103/PhysRevD.87.124032
[arXiv:1210.0731 [hep-th]].
%132 citations counted in INSPIRE as of 26 Jun 2024

%\cite{Duval:2014uva}
\bibitem{Duval:2014uva}
C.~Duval, G.~W.~Gibbons and P.~A.~Horvathy,
``Conformal Carroll groups and BMS symmetry,''
Class. Quant. Grav. \textbf{31}, 092001 (2014)
doi:10.1088/0264-9381/31/9/092001
[arXiv:1402.5894 [gr-qc]].
%318 citations counted in INSPIRE as of 31 Mar 2025


%\cite{Bagchi:2019xfx}
\bibitem{Bagchi:2019xfx}
A.~Bagchi, A.~Mehra and P.~Nandi,
``Field Theories with Conformal Carrollian Symmetry,''
JHEP \textbf{05} (2019), 108
doi:10.1007/JHEP05(2019)108
[arXiv:1901.10147 [hep-th]].
%98 citations counted in INSPIRE as of 26 Jun 2024


%\cite{Henneaux:2021yzg}
\bibitem{Henneaux:2021yzg}
M.~Henneaux and P.~Salgado-Rebolledo,
``Carroll contractions of Lorentz-invariant theories,''
JHEP \textbf{11} (2021), 180
doi:10.1007/JHEP11(2021)180
[arXiv:2109.06708 [hep-th]].
%98 citations counted in INSPIRE as of 26 Jun 2024


%\cite{deBoer:2021jej}
\bibitem{deBoer:2021jej}
J.~de Boer, J.~Hartong, N.~A.~Obers, W.~Sybesma and S.~Vandoren,
``Carroll Symmetry, Dark Energy and Inflation,''
Front. in Phys. \textbf{10} (2022), 810405
doi:10.3389/fphy.2022.810405
[arXiv:2110.02319 [hep-th]].
%99 citations counted in INSPIRE as of 26 Jun 2024

%\cite{Chen:2023pqf}
\bibitem{Chen:2023pqf}
B.~Chen, R.~Liu, H.~Sun and Y.~f.~Zheng,
``Constructing Carrollian field theories from null reduction,''
JHEP \textbf{11} (2023), 170
doi:10.1007/JHEP11(2023)170
[arXiv:2301.06011 [hep-th]].
%23 citations counted in INSPIRE as of 26 Jun 2024




%\cite{Salzer:2023jqv}
\bibitem{Salzer:2023jqv}
J.~Salzer,
``An embedding space approach to Carrollian CFT correlators for flat space holography,''
JHEP \textbf{10}, 084 (2023)
doi:10.1007/JHEP10(2023)084
[arXiv:2304.08292 [hep-th]].
%26 citations counted in INSPIRE as of 11 Jun 2024

%\cite{Saha:2023abr}
\bibitem{Saha:2023abr}
A.~Saha,
``w$_{1+\infty}$ and Carrollian holography,''
JHEP \textbf{05}, 145 (2024)
doi:10.1007/JHEP05(2024)145
[arXiv:2308.03673 [hep-th]].
%14 citations counted in INSPIRE as of 26 Jun 2024

%\cite{Nguyen:2023vfz}
\bibitem{Nguyen:2023vfz}
K.~Nguyen and P.~West,
``Carrollian Conformal Fields and Flat Holography,''
Universe \textbf{9}, no.9, 385 (2023)
doi:10.3390/universe9090385
[arXiv:2305.02884 [hep-th]].
%22 citations counted in INSPIRE as of 10 Jul 2024

%\cite{Nguyen:2023miw}
\bibitem{Nguyen:2023miw}
K.~Nguyen,
``Carrollian conformal correlators and massless scattering amplitudes,''
JHEP \textbf{01}, 076 (2024)
doi:10.1007/JHEP01(2024)076
[arXiv:2311.09869 [hep-th]].
%8 citations counted in INSPIRE as of 15 May 2024

%\cite{Bagchi:2023cen}
\bibitem{Bagchi:2023cen}
A.~Bagchi, P.~Dhivakar and S.~Dutta,
``Holography in Flat Spacetimes: the case for Carroll,''
[arXiv:2311.11246 [hep-th]].
%16 citations counted in INSPIRE as of 11 Jun 2024

%\cite{Mason:2023mti}
\bibitem{Mason:2023mti}
L.~Mason, R.~Ruzziconi and A.~Yelleshpur Srikant,
``Carrollian Amplitudes and Celestial Symmetries,''
[arXiv:2312.10138 [hep-th]].
%10 citations counted in INSPIRE as of 07 May 2024

%\cite{Liu:2024nfc}
\bibitem{Liu:2024nfc}
W.~B.~Liu, J.~Long and X.~Q.~Ye,
``Feynman rules and loop structure of Carrollian amplitudes,''
JHEP \textbf{05}, 213 (2024)
doi:10.1007/JHEP05(2024)213
[arXiv:2402.04120 [hep-th]].
%3 citations counted in INSPIRE as of 11 Jun 2024

%\cite{Have:2024dff}
\bibitem{Have:2024dff}
E.~Have, K.~Nguyen, S.~Prohazka and J.~Salzer,
``Massive carrollian fields at timelike infinity,''
[arXiv:2402.05190 [hep-th]].
%6 citations counted in INSPIRE as of 11 Jun 2024

%\cite{Stieberger:2024shv}
\bibitem{Stieberger:2024shv}
S.~Stieberger, T.~R.~Taylor and B.~Zhu,
``Carrollian Amplitudes from Strings,''
JHEP \textbf{04}, 127 (2024)
doi:10.1007/JHEP04(2024)127
[arXiv:2402.14062 [hep-th]].
%3 citations counted in INSPIRE as of 10 May 2024

%\cite{Adamo:2024mqn}
\bibitem{Adamo:2024mqn}
T.~Adamo, W.~Bu, P.~Tourkine and B.~Zhu,
``Eikonal amplitudes on the celestial sphere,''
[arXiv:2405.15594 [hep-th]].
%0 citations counted in INSPIRE as of 11 Jun 2024


%\cite{Banerjee:2024hvb,Liu:2024llk,Ruzziconi:2024kzo,Chakrabortty:2024bvm}
\bibitem{Banerjee:2024hvb}
S.~Banerjee, R.~Basu and S.~Atul Bhatkar,
``Light transformation: a celestial and Carrollian perspective,''
JHEP \textbf{12}, 122 (2024)
doi:10.1007/JHEP12(2024)122
[arXiv:2407.08379 [hep-th]].
%8 citations counted in INSPIRE as of 25 Mar 2025


%\cite{Liu:2024llk}
\bibitem{Liu:2024llk}
W.~B.~Liu, J.~Long, H.~Y.~Xiao and J.~L.~Yang,
``On the definition of Carrollian amplitudes in general dimensions,''
JHEP \textbf{11}, 027 (2024)
doi:10.1007/JHEP11(2024)027
[arXiv:2407.20816 [hep-th]].
%5 citations counted in INSPIRE as of 25 Mar 2025

%\cite{Ruzziconi:2024zkr}
\bibitem{Ruzziconi:2024zkr}
R.~Ruzziconi, S.~Stieberger, T.~R.~Taylor and B.~Zhu,
``Differential equations for Carrollian amplitudes,''
JHEP \textbf{09}, 149 (2024)
doi:10.1007/JHEP09(2024)149
[arXiv:2407.04789 [hep-th]].
%12 citations counted in INSPIRE as of 26 Mar 2025


%\cite{Ruzziconi:2024kzo}
\bibitem{Ruzziconi:2024kzo}
R.~Ruzziconi and A.~Saha,
``Holographic Carrollian currents for massless scattering,''
JHEP \textbf{01}, 169 (2025)
doi:10.1007/JHEP01(2025)169
[arXiv:2411.04902 [hep-th]].
%3 citations counted in INSPIRE as of 25 Mar 2025


%\cite{Chakrabortty:2024bvm}
\bibitem{Chakrabortty:2024bvm}
S.~Chakrabortty, S.~Hegde and A.~Maurya,
``Differential Representation for Carrollian Correlators,''
[arXiv:2411.09641 [hep-th]].
%0 citations counted in INSPIRE as of 25 Mar 2025

%\cite{Nguyen:2025sqk}
\bibitem{Nguyen:2025sqk}
K.~Nguyen and J.~Salzer,
``Operator Product Expansion in Carrollian CFT,''
[arXiv:2503.15607 [hep-th]].
%0 citations counted in INSPIRE as of 25 Mar 2025

%\cite{Maldacena:1997re}
\bibitem{Maldacena:1997re}
J.~M.~Maldacena,
``The Large N limit of superconformal field theories and supergravity,''
Adv. Theor. Math. Phys. \textbf{2}, 231-252 (1998)
doi:10.4310/ATMP.1998.v2.n2.a1
[arXiv:hep-th/9711200 [hep-th]].
%20644 citations counted in INSPIRE as of 25 Mar 2025


%\cite{Witten:1998qj}
\bibitem{Witten:1998qj}
E.~Witten,
``Anti-de Sitter space and holography,''
Adv. Theor. Math. Phys. \textbf{2}, 253-291 (1998)
doi:10.4310/ATMP.1998.v2.n2.a2
[arXiv:hep-th/9802150 [hep-th]].
%13175 citations counted in INSPIRE as of 25 Mar 2025

%\cite{Aharony:1999ti}
\bibitem{Aharony:1999ti}
O.~Aharony, S.~S.~Gubser, J.~M.~Maldacena, H.~Ooguri and Y.~Oz,
``Large N field theories, string theory and gravity,''
Phys. Rept. \textbf{323}, 183-386 (2000)
doi:10.1016/S0370-1573(99)00083-6
[arXiv:hep-th/9905111 [hep-th]].
%5900 citations counted in INSPIRE as of 25 Mar 2025

%\cite{Susskind:1998vk}
\bibitem{Susskind:1998vk}
L.~Susskind,
``Holography in the flat space limit,''
AIP Conf. Proc. \textbf{493}, no.1, 98-112 (1999)
doi:10.1063/1.1301570
[arXiv:hep-th/9901079 [hep-th]].
%258 citations counted in INSPIRE as of 25 Mar 2025

%\cite{Polchinski:1999ry}
\bibitem{Polchinski:1999ry}
J.~Polchinski,
``S matrices from AdS space-time,''
[arXiv:hep-th/9901076 [hep-th]].
%247 citations counted in INSPIRE as of 25 Mar 2025

%\cite{Giddings:1999jq}
\bibitem{Giddings:1999jq}
S.~B.~Giddings,
``Flat space scattering and bulk locality in the AdS / CFT correspondence,''
Phys. Rev. D \textbf{61}, 106008 (2000)
doi:10.1103/PhysRevD.61.106008
[arXiv:hep-th/9907129 [hep-th]].
%155 citations counted in INSPIRE as of 25 Mar 2025

%\cite{Marotta:2024sce}
\bibitem{Marotta:2024sce}
R.~Marotta, K.~Skenderis and M.~Verma,
``Flat space spinning massive amplitudes from momentum space CFT,''
JHEP \textbf{08}, 226 (2024)
doi:10.1007/JHEP08(2024)226
[arXiv:2406.06447 [hep-th]].
%8 citations counted in INSPIRE as of 25 Mar 2025

%\cite{Gary:2009ae}
\bibitem{Gary:2009ae}
M.~Gary, S.~B.~Giddings and J.~Penedones,
``Local bulk S-matrix elements and CFT singularities,''
Phys. Rev. D \textbf{80} (2009), 085005
doi:10.1103/PhysRevD.80.085005
[arXiv:0903.4437 [hep-th]].
%184 citations counted in INSPIRE as of 06 Feb 2025

%\cite{Penedones:2010ue}
\bibitem{Penedones:2010ue}
J.~Penedones,
``Writing CFT correlation functions as AdS scattering amplitudes,''
JHEP \textbf{03}, 025 (2011)
doi:10.1007/JHEP03(2011)025
[arXiv:1011.1485 [hep-th]].
%583 citations counted in INSPIRE as of 25 Mar 2025


%\cite{Fitzpatrick:2011hu}
\bibitem{Fitzpatrick:2011hu}
A.~L.~Fitzpatrick and J.~Kaplan,
``Analyticity and the Holographic S-Matrix,''
JHEP \textbf{10}, 127 (2012)
doi:10.1007/JHEP10(2012)127
[arXiv:1111.6972 [hep-th]].
%208 citations counted in INSPIRE as of 25 Mar 2025


%\cite{Alday:2017vkk}
\bibitem{Alday:2017vkk}
L.~F.~Alday and S.~Caron-Huot,
``Gravitational S-matrix from CFT dispersion relations,''
JHEP \textbf{12}, 017 (2018)
doi:10.1007/JHEP12(2018)017
[arXiv:1711.02031 [hep-th]].
%181 citations counted in INSPIRE as of 25 Mar 2025

%\cite{Hijano:2019qmi}
\bibitem{Hijano:2019qmi}
E.~Hijano,
``Flat space physics from AdS/CFT,''
JHEP \textbf{07}, 132 (2019)
doi:10.1007/JHEP07(2019)132
[arXiv:1905.02729 [hep-th]].
%82 citations counted in INSPIRE as of 25 Mar 2025

%\cite{Li:2021snj}
\bibitem{Li:2021snj}
Y.~Z.~Li,
``Notes on flat-space limit of AdS/CFT,''
JHEP \textbf{09}, 027 (2021)
doi:10.1007/JHEP09(2021)027
[arXiv:2106.04606 [hep-th]].
%60 citations counted in INSPIRE as of 25 Mar 2025


%\cite{Lam:2017ofc}
\bibitem{Lam:2017ofc}
H.~T.~Lam and S.~H.~Shao,
``Conformal Basis, Optical Theorem, and the Bulk Point Singularity,''
Phys. Rev. D \textbf{98} (2018) no.2, 025020
doi:10.1103/PhysRevD.98.025020
[arXiv:1711.06138 [hep-th]].
%84 citations counted in INSPIRE as of 06 Feb 2025

%\cite{deGioia:2022fcn}
\bibitem{deGioia:2022fcn}
L.~P.~de Gioia and A.~M.~Raclariu,
``Eikonal approximation in celestial CFT,''
JHEP \textbf{03}, 030 (2023)
doi:10.1007/JHEP03(2023)030
[arXiv:2206.10547 [hep-th]].
%47 citations counted in INSPIRE as of 25 Mar 2025


%\cite{deGioia:2024yne}
\bibitem{deGioia:2024yne}
L.~P.~de Gioia and A.~M.~Raclariu,
``Celestial amplitudes from conformal correlators with bulk-point kinematics,''
[arXiv:2405.07972 [hep-th]].
%8 citations counted in INSPIRE as of 25 Mar 2025

%\cite{Bagchi:2023fbj}
\bibitem{Bagchi:2023fbj}
A.~Bagchi, P.~Dhivakar and S.~Dutta,
``AdS Witten diagrams to Carrollian correlators,''
JHEP \textbf{04}, 135 (2023)
doi:10.1007/JHEP04(2023)135
[arXiv:2303.07388 [hep-th]].
%56 citations counted in INSPIRE as of 25 Mar 2025

%\cite{Kawai:1985xq}
\bibitem{Kawai:1985xq}
H.~Kawai, D.~C.~Lewellen and S.~H.~H.~Tye,
``A Relation Between Tree Amplitudes of Closed and Open Strings,''
Nucl. Phys. B \textbf{269} (1986), 1-23
doi:10.1016/0550-3213(86)90362-7
%1277 citations counted in INSPIRE as of 20 Dec 2024

%\cite{Bern:2008qj}
\bibitem{Bern:2008qj}
Z.~Bern, J.~J.~M.~Carrasco and H.~Johansson,
%``New Relations for Gauge-Theory Amplitudes,''
Phys. Rev. D \textbf{78} (2008), 085011
doi:10.1103/PhysRevD.78.085011
[arXiv:0805.3993 [hep-ph]].
%1293 citations counted in INSPIRE as of 20 Dec 2024

%\cite{Bern:2019prr}
\bibitem{Bern:2019prr}
Z.~Bern, J.~J.~Carrasco, M.~Chiodaroli, H.~Johansson and R.~Roiban,
``The duality between color and kinematics and its applications,''
J. Phys. A \textbf{57}, no.33, 333002 (2024)
doi:10.1088/1751-8121/ad5fd0
[arXiv:1909.01358 [hep-th]].
%514 citations counted in INSPIRE as of 10 Apr 2025

%\cite{Adamo:2022dcm}
\bibitem{Adamo:2022dcm}
T.~Adamo, J.~J.~M.~Carrasco, M.~Carrillo-Gonz\'alez, M.~Chiodaroli, H.~Elvang, H.~Johansson, D.~O'Connell, R.~Roiban and O.~Schlotterer,
``Snowmass White Paper: the Double Copy and its Applications,''
[arXiv:2204.06547 [hep-th]].
%121 citations counted in INSPIRE as of 10 Apr 2025

%\cite{Arkani-Hamed:2020gyp}
\bibitem{Arkani-Hamed:2020gyp}
N.~Arkani-Hamed, M.~Pate, A.~M.~Raclariu and A.~Strominger,
``Celestial amplitudes from UV to IR,''
JHEP \textbf{08}, 062 (2021)
doi:10.1007/JHEP08(2021)062
[arXiv:2012.04208 [hep-th]].
%132 citations counted in INSPIRE as of 10 Apr 2025


%\cite{Stieberger:2018onx,Fan:2020xjj}
\bibitem{Stieberger:2018onx}
S.~Stieberger and T.~R.~Taylor,
``Symmetries of Celestial Amplitudes,''
Phys. Lett. B \textbf{793}, 141-143 (2019)
doi:10.1016/j.physletb.2019.03.063
[arXiv:1812.01080 [hep-th]].
%110 citations counted in INSPIRE as of 25 Mar 2025


%\cite{Fan:2020xjj}
\bibitem{Fan:2020xjj}
W.~Fan, A.~Fotopoulos, S.~Stieberger and T.~R.~Taylor,
``On Sugawara construction on Celestial Sphere,''
JHEP \textbf{09}, 139 (2020)
doi:10.1007/JHEP09(2020)139
[arXiv:2005.10666 [hep-th]].
%38 citations counted in INSPIRE as of 25 Mar 2025


%\cite{Casali:2020vuy}
\bibitem{Casali:2020vuy}
E.~Casali and A.~Puhm,
``Double Copy for Celestial Amplitudes,''
Phys. Rev. Lett. \textbf{126} (2021) no.10, 101602
doi:10.1103/PhysRevLett.126.101602
[arXiv:2007.15027 [hep-th]].
%70 citations counted in INSPIRE as of 21 Dec 2024

%\cite{Casali:2020uvr}
\bibitem{Casali:2020uvr}
E.~Casali and A.~Sharma,
``Celestial double copy from the worldsheet,''
JHEP \textbf{05} (2021), 157
doi:10.1007/JHEP05(2021)157
[arXiv:2011.10052 [hep-th]].
%46 citations counted in INSPIRE as of 21 Dec 2024

%\cite{Surubaru:2025qhs}
\bibitem{Surubaru:2025qhs}
I.~Surubaru and B.~Zhu,
``Conformal blocks from celestial graviton amplitudes,''
[arXiv:2501.05805 [hep-th]].
%0 citations counted in INSPIRE as of 25 Mar 2025

%\cite{He:2019jjk}
\bibitem{He:2019jjk}
T.~He and P.~Mitra,
``Asymptotic symmetries and Weinberg\textquoteright{}s soft photon theorem in Mink$_{d+2}$,''
JHEP \textbf{10}, 213 (2019)
doi:10.1007/JHEP10(2019)213
[arXiv:1903.02608 [hep-th]].
%46 citations counted in INSPIRE as of 24 Feb 2025

%\cite{Gonzo:2020xza}
\bibitem{Gonzo:2020xza}
R.~Gonzo and A.~Pokraka,
``Light-ray operators, detectors and gravitational event shapes,''
JHEP \textbf{05}, 015 (2021)
doi:10.1007/JHEP05(2021)015
[arXiv:2012.01406 [hep-th]].
%29 citations counted in INSPIRE as of 24 Feb 2025

%\cite{Barnich:2012aw}
\bibitem{Barnich:2012aw}
G.~Barnich, A.~Gomberoff and H.~A.~Gonzalez,
``The Flat limit of three dimensional asymptotically anti-de Sitter spacetimes,''
Phys. Rev. D \textbf{86}, 024020 (2012)
doi:10.1103/PhysRevD.86.024020
[arXiv:1204.3288 [gr-qc]].
%250 citations counted in INSPIRE as of 24 Feb 2025



%\cite{Duval:2014uoa,Henneaux:2021yzg,deBoer:2021jej,Hansen:2021fxi,Baiguera:2022lsw}
\bibitem{Duval:2014uoa}
C.~Duval, G.~W.~Gibbons, P.~A.~Horvathy and P.~M.~Zhang,
``Carroll versus Newton and Galilei: two dual non-Einsteinian concepts of time,''
Class. Quant. Grav. \textbf{31}, 085016 (2014)
doi:10.1088/0264-9381/31/8/085016
[arXiv:1402.0657 [gr-qc]].
%320 citations counted in INSPIRE as of 09 Apr 2025





%\cite{Hansen:2021fxi}
\bibitem{Hansen:2021fxi}
D.~Hansen, N.~A.~Obers, G.~Oling and B.~T.~S\o{}gaard,
``Carroll Expansion of General Relativity,''
SciPost Phys. \textbf{13}, no.3, 055 (2022)
doi:10.21468/SciPostPhys.13.3.055
[arXiv:2112.12684 [hep-th]].
%84 citations counted in INSPIRE as of 09 Apr 2025


%\cite{Baiguera:2022lsw}
\bibitem{Baiguera:2022lsw}
S.~Baiguera, G.~Oling, W.~Sybesma and B.~T.~S\o{}gaard,
``Conformal Carroll scalars with boosts,''
SciPost Phys. \textbf{14}, no.4, 086 (2023)
doi:10.21468/SciPostPhys.14.4.086
[arXiv:2207.03468 [hep-th]].
%61 citations counted in INSPIRE as of 27 Mar 2025



%\cite{Ben-Shahar:2021zww}
\bibitem{Ben-Shahar:2021zww}
M.~Ben-Shahar and H.~Johansson,
``Off-shell color-kinematics duality for Chern-Simons,''
JHEP \textbf{08} (2022), 035
doi:10.1007/JHEP08(2022)035
[arXiv:2112.11452 [hep-th]].
%57 citations counted in INSPIRE as of 14 Mar 2025





%\cite{Barker:1966zz,Huggins:1987ea}
\bibitem{Barker:1966zz}
B.~M.~Barker, S.~N.~Gupta and R.~D.~Haracz,
``One-Graviton Exchange Interaction of Elementary Particles,''
Phys. Rev. \textbf{149}, 1027-1032 (1966)
doi:10.1103/PhysRev.149.1027
%54 citations counted in INSPIRE as of 13 Mar 2025

%\cite{Huggins:1987ea}
\bibitem{Huggins:1987ea}
S.~R.~Huggins and D.~J.~Toms,
``One Graviton Exchange Interaction of Nonminimally Coupled Scalar Fields,''
Class. Quant. Grav. \textbf{4}, 1509 (1987)
doi:10.1088/0264-9381/4/6/010
%20 citations counted in INSPIRE as of 13 Mar 2025





%\cite{DHoker:1999mqo}
\bibitem{DHoker:1999mqo}
E.~D'Hoker, D.~Z.~Freedman and L.~Rastelli,
``AdS / CFT four point functions: How to succeed at z integrals without really trying,''
Nucl. Phys. B \textbf{562}, 395-411 (1999)
doi:10.1016/S0550-3213(99)00526-X
[arXiv:hep-th/9905049 [hep-th]].
%195 citations counted in INSPIRE as of 21 Feb 2025

%\cite{Bissi:2022mrs}
\bibitem{Bissi:2022mrs}
A.~Bissi, A.~Sinha and X.~Zhou,
``Selected topics in analytic conformal bootstrap: A guided journey,''
Phys. Rept. \textbf{991}, 1-89 (2022)
doi:10.1016/j.physrep.2022.09.004
[arXiv:2202.08475 [hep-th]].
%95 citations counted in INSPIRE as of 21 Feb 2025





\bibitem{Duffin}
D. Simmons-Duffin, ``TASI Lectures on Conformal Field Theory
in Lorentzian Signature,'' (2019).


%\cite{Zhang:2008jy}
\bibitem{Zhang:2008jy}
H.~H.~Zhang, K.~X.~Feng, S.~W.~Qiu, A.~Zhao and X.~S.~Li,
``On analytic formulas of Feynman propagators in position space,''
Chin. Phys. C \textbf{34}, 1576-1582 (2010)
doi:10.1088/1674-1137/34/10/005
[arXiv:0811.1261 [math-ph]].
%36 citations counted in INSPIRE as of 24 Mar 2025

%\cite{DHoker:1999bve}
\bibitem{DHoker:1999bve}
E.~D'Hoker, D.~Z.~Freedman, S.~D.~Mathur, A.~Matusis and L.~Rastelli,
``Graviton and gauge boson propagators in AdS(d+1),''
Nucl. Phys. B \textbf{562}, 330-352 (1999)
doi:10.1016/S0550-3213(99)00524-6
[arXiv:hep-th/9902042 [hep-th]].
%162 citations counted in INSPIRE as of 02 Apr 2025


%\cite{Denner:1991qq,Usyukina:1992jd}
\bibitem{Denner:1991qq}
A.~Denner, U.~Nierste and R.~Scharf,
``A Compact expression for the scalar one loop four point function,''
Nucl. Phys. B \textbf{367}, 637-656 (1991)
doi:10.1016/0550-3213(91)90011-L
%273 citations counted in INSPIRE as of 24 Mar 2025

%\cite{Usyukina:1992jd}
\bibitem{Usyukina:1992jd}
N.~I.~Usyukina and A.~I.~Davydychev,
``An Approach to the evaluation of three and four point ladder diagrams,''
Phys. Lett. B \textbf{298}, 363-370 (1993)
doi:10.1016/0370-2693(93)91834-A
%273 citations counted in INSPIRE as of 24 Mar 2025


%\cite{Bourjaily:2020wvq}
\bibitem{Bourjaily:2020wvq}
J.~L.~Bourjaily, H.~Hannesdottir, A.~J.~McLeod, M.~D.~Schwartz and C.~Vergu,
``Sequential Discontinuities of Feynman Integrals and the Monodromy Group,''
JHEP \textbf{01}, 205 (2021)
doi:10.1007/JHEP01(2021)205
[arXiv:2007.13747 [hep-th]].
%64 citations counted in INSPIRE as of 24 Mar 2025

%\cite{Maldacena:2015iua}
\bibitem{Maldacena:2015iua}
J.~Maldacena, D.~Simmons-Duffin and A.~Zhiboedov,
``Looking for a bulk point,''
JHEP \textbf{01}, 013 (2017)
doi:10.1007/JHEP01(2017)013
[arXiv:1509.03612 [hep-th]].
%282 citations counted in INSPIRE as of 24 Mar 2025



%\cite{Rastelli:2019gtj}
\bibitem{Rastelli:2019gtj}
L.~Rastelli, K.~Roumpedakis and X.~Zhou,
``$\mathbf{AdS_3\times S^3}$ Tree-Level Correlators: Hidden Six-Dimensional Conformal Symmetry,''
JHEP \textbf{10} (2019), 140
doi:10.1007/JHEP10(2019)140
[arXiv:1905.11983 [hep-th]].
%85 citations counted in INSPIRE as of 11 Mar 2025






%\cite{Cornalba:2006xk}
\bibitem{Cornalba:2006xk}
L.~Cornalba, M.~S.~Costa, J.~Penedones and R.~Schiappa,
``Eikonal Approximation in AdS/CFT: From Shock Waves to Four-Point Functions,''
JHEP \textbf{08}, 019 (2007)
doi:10.1088/1126-6708/2007/08/019
[arXiv:hep-th/0611122 [hep-th]].
%166 citations counted in INSPIRE as of 27 Feb 2025


\bibitem{DiVecchia:2023frv}
P.~Di Vecchia, C.~Heissenberg, R.~Russo and G.~Veneziano,
``The gravitational eikonal: From particle, string and brane collisions to black-hole encounters,''
Phys. Rept. \textbf{1083}, 1-169 (2024)
doi:10.1016/j.physrep.2024.06.002
[arXiv:2306.16488 [hep-th]].
%85 citations counted in INSPIRE as of 07 Mar 2025

%\cite{tHooft:1987vrq}
\bibitem{tHooft:1987vrq}
G.~'t Hooft,
``Graviton Dominance in Ultrahigh-Energy Scattering,''
Phys. Lett. B \textbf{198}, 61-63 (1987)
doi:10.1016/0370-2693(87)90159-6
%534 citations counted in INSPIRE as of 01 Apr 2025

%\cite{Keski-Vakkuri:1998gmz}
\bibitem{Keski-Vakkuri:1998gmz}
E.~Keski-Vakkuri,
``Bulk and boundary dynamics in BTZ black holes,''
Phys. Rev. D \textbf{59} (1999), 104001
doi:10.1103/PhysRevD.59.104001
[arXiv:hep-th/9808037 [hep-th]].
%147 citations counted in INSPIRE as of 09 Apr 2025

%\cite{deBoer:2023fnj}
\bibitem{deBoer:2023fnj}
J.~de Boer, J.~Hartong, N.~A.~Obers, W.~Sybesma and S.~Vandoren,
``Carroll stories,''
JHEP \textbf{09}, 148 (2023)
doi:10.1007/JHEP09(2023)148
[arXiv:2307.06827 [hep-th]].
%80 citations counted in INSPIRE as of 27 Mar 2025


%\cite{Chen:2024voz}
\bibitem{Chen:2024voz}
B.~Chen, H.~Sun and Y.~f.~Zheng,
``Quantization of Carrollian conformal scalar theories,''
Phys. Rev. D \textbf{110}, no.12, 125010 (2024)
doi:10.1103/PhysRevD.110.125010
[arXiv:2406.17451 [hep-th]].
%6 citations counted in INSPIRE as of 27 Mar 2025




%\cite{Cotler:2024xhb}
\bibitem{Cotler:2024xhb}
J.~Cotler, K.~Jensen, S.~Prohazka, A.~Raz, M.~Riegler and J.~Salzer,
``Quantizing Carrollian field theories,''
JHEP \textbf{10}, 049 (2024)
doi:10.1007/JHEP10(2024)049
[arXiv:2407.11971 [hep-th]].
%10 citations counted in INSPIRE as of 27 Mar 2025

%\cite{Kraus:2024gso}
\bibitem{Kraus:2024gso}
P.~Kraus and R.~M.~Myers,
``Carrollian partition functions and the flat limit of AdS,''
JHEP \textbf{01}, 183 (2025)
doi:10.1007/JHEP01(2025)183
[arXiv:2407.13668 [hep-th]].
%13 citations counted in INSPIRE as of 27 Mar 2025

%\cite{Kraus:2025wgi}
\bibitem{Kraus:2025wgi}
P.~Kraus and R.~M.~Myers,
``Carrollian Partition Function for Bulk Yang-Mills Theory,''
[arXiv:2503.00916 [hep-th]].
%2 citations counted in INSPIRE as of 27 Mar 2025



%\cite{Figueroa-OFarrill:2023qty}
\bibitem{Figueroa-OFarrill:2023qty}
J.~Figueroa-O'Farrill, A.~P\'erez and S.~Prohazka,
``Quantum Carroll/fracton particles,''
JHEP \textbf{10}, 041 (2023)
doi:10.1007/JHEP10(2023)041
[arXiv:2307.05674 [hep-th]].
%36 citations counted in INSPIRE as of 27 Mar 2025

%\cite{Ciambelli:2018wre}
\bibitem{Ciambelli:2018wre}
L.~Ciambelli, C.~Marteau, A.~C.~Petkou, P.~M.~Petropoulos and K.~Siampos,
``Flat holography and Carrollian fluids,''
JHEP \textbf{07}, 165 (2018)
doi:10.1007/JHEP07(2018)165
[arXiv:1802.06809 [hep-th]].
%161 citations counted in INSPIRE as of 31 Mar 2025

%\cite{Ciambelli:2019lap}
\bibitem{Ciambelli:2019lap}
L.~Ciambelli, R.~G.~Leigh, C.~Marteau and P.~M.~Petropoulos,
``Carroll Structures, Null Geometry and Conformal Isometries,''
Phys. Rev. D \textbf{100}, no.4, 046010 (2019)
doi:10.1103/PhysRevD.100.046010
[arXiv:1905.02221 [hep-th]].
%150 citations counted in INSPIRE as of 31 Mar 2025

%\cite{Cotler:2024cia}
\bibitem{Cotler:2024cia}
J.~Cotler, K.~Jensen, S.~Prohazka, M.~Riegler and J.~Salzer,
``Soft gravitons in three dimensions,''
[arXiv:2411.13633 [hep-th]].
%4 citations counted in INSPIRE as of 27 Mar 2025


%\cite{Poulias:2025eck}
\bibitem{Poulias:2025eck}
G.~Poulias and S.~Vandoren,
``On Carroll partition functions and flat space holography,''
[arXiv:2503.20615 [hep-th]].
%0 citations counted in INSPIRE as of 27 Mar 2025

%\cite{Bagchi:2016geg,Bagchi:2017cpu}
\bibitem{Bagchi:2016geg}
A.~Bagchi, M.~Gary and Zodinmawia,
``Bondi-Metzner-Sachs bootstrap,''
Phys. Rev. D \textbf{96}, no.2, 025007 (2017)
doi:10.1103/PhysRevD.96.025007
[arXiv:1612.01730 [hep-th]].
%59 citations counted in INSPIRE as of 01 Apr 2025

%\cite{Bagchi:2017cpu}
\bibitem{Bagchi:2017cpu}
A.~Bagchi, M.~Gary and Zodinmawia,
``The nuts and bolts of the BMS Bootstrap,''
Class. Quant. Grav. \textbf{34}, no.17, 174002 (2017)
doi:10.1088/1361-6382/aa8003
[arXiv:1705.05890 [hep-th]].
%41 citations counted in INSPIRE as of 01 Apr 2025

%\cite{Chen:2020vvn,Chen:2022cpx,Chen:2022jhx}
\bibitem{Chen:2020vvn}
B.~Chen, P.~X.~Hao, R.~Liu and Z.~F.~Yu,
``On Galilean conformal bootstrap,''
JHEP \textbf{06}, 112 (2021)
doi:10.1007/JHEP06(2021)112
[arXiv:2011.11092 [hep-th]].
%24 citations counted in INSPIRE as of 11 Apr 2025



%\cite{Chen:2022cpx}
\bibitem{Chen:2022cpx}
B.~Chen and R.~Liu,
``The shadow formalism of Galilean CFT$_{2}$,''
JHEP \textbf{05}, 224 (2023)
doi:10.1007/JHEP05(2023)224
[arXiv:2203.10490 [hep-th]].
%19 citations counted in INSPIRE as of 11 Apr 2025

%\cite{Chen:2022jhx}
\bibitem{Chen:2022jhx}
B.~Chen, P.~x.~Hao, R.~Liu and Z.~f.~Yu,
``On Galilean conformal bootstrap. Part II. \ensuremath{\xi} = 0 sector,''
JHEP \textbf{12}, 019 (2022)
doi:10.1007/JHEP12(2022)019
[arXiv:2207.01474 [hep-th]].
%16 citations counted in INSPIRE as of 11 Apr 2025


%\cite{Hijano:2015zsa}
\bibitem{Hijano:2015zsa}
E.~Hijano, P.~Kraus, E.~Perlmutter and R.~Snively,
``Witten Diagrams Revisited: The AdS Geometry of Conformal Blocks,''
JHEP \textbf{01}, 146 (2016)
doi:10.1007/JHEP01(2016)146
[arXiv:1508.00501 [hep-th]].
%268 citations counted in INSPIRE as of 01 Apr 2025

%\cite{Hijano:2015qja}
\bibitem{Hijano:2015qja}
E.~Hijano, P.~Kraus, E.~Perlmutter and R.~Snively,
``Semiclassical Virasoro blocks from AdS$_{3}$ gravity,''
JHEP \textbf{12}, 077 (2015)
doi:10.1007/JHEP12(2015)077
[arXiv:1508.04987 [hep-th]].
%143 citations counted in INSPIRE as of 01 Apr 2025


%\cite{Fan:2022vbz}
\bibitem{Fan:2022vbz}
W.~Fan, A.~Fotopoulos, S.~Stieberger, T.~R.~Taylor and B.~Zhu,
``Elements of celestial conformal field theory,''
JHEP \textbf{08}, 213 (2022)
doi:10.1007/JHEP08(2022)213
[arXiv:2202.08288 [hep-th]].
%49 citations counted in INSPIRE as of 01 Apr 2025

%\cite{Yu:2022bcp}
\bibitem{Yu:2022bcp}
Z.~f.~Yu and B.~Chen,
``Free field realization of the BMS Ising model,''
JHEP \textbf{08}, 116 (2023)
doi:10.1007/JHEP08(2023)116
[arXiv:2211.06926 [hep-th]].
%24 citations counted in INSPIRE as of 09 Apr 2025


%\cite{Hao:2022xhq}
\bibitem{Hao:2022xhq}
P.~X.~Hao, W.~Song, Z.~Xiao and X.~Xie,
``BMS-invariant free fermion models,''
Phys. Rev. D \textbf{109}, no.2, 025002 (2024)
doi:10.1103/PhysRevD.109.025002
[arXiv:2211.06927 [hep-th]].
%24 citations counted in INSPIRE as of 11 Apr 2025

\end{thebibliography}
\end{document}